\documentclass[twocolumn,english,aps,pra,showpacs]{revtex4}
\usepackage[T1]{fontenc}
\usepackage[latin9]{inputenc}
\usepackage{amsthm}
\usepackage{amsmath}
\usepackage{amssymb}
\usepackage{graphicx}
\usepackage{esint}

\makeatletter
\@ifundefined{textcolor}{}
{%
 \definecolor{BLACK}{gray}{0}
 \definecolor{WHITE}{gray}{1}
 \definecolor{RED}{rgb}{1,0,0}
 \definecolor{GREEN}{rgb}{0,1,0}
 \definecolor{BLUE}{rgb}{0,0,1}
 \definecolor{CYAN}{cmyk}{1,0,0,0}
 \definecolor{MAGENTA}{cmyk}{0,1,0,0}
 \definecolor{YELLOW}{cmyk}{0,0,1,0}
}
  \theoremstyle{plain}
  \ifx\thechapter\undefined
    \newtheorem{prop}{\protect\propositionname}
  \else
    \newtheorem{prop}{\protect\propositionname}[chapter]
  \fi

\usepackage{subfigure}\usepackage{times}\usepackage{mathrsfs}\usepackage{longtable}\usepackage{latexsym}\usepackage{bm}

\usepackage{braket}\newcommand{\be }{\begin {equation}} \newcommand{\ee }{\end {equation}}

\usepackage{babel}
\providecommand{\propositionname}{Proposition}

\makeatother

\usepackage{babel}
  \providecommand{\propositionname}{Proposition}

\begin{document}

\title{Two-qubit correlations revisited: average mutual information, relevant
(and useful) observables and an application to remote state preparation}

\author{Paolo Giorda$^{1}$ }

\email{magpaolo16@gmail.com}

\author{Michele Allegra$^{2}$}

\email{mallegra@sissa.it}

\affiliation{$^{1}$Consorzio Nazionale Interuniversitario per le Scienze Fisiche
della Materia (CNISM), I-20133 Milan, Italy}

\affiliation{$^{2}$Scuola Internazionale Superiore di Studi Avanzati (SISSA),
I-34136 Trieste, Italy }
\begin{abstract}
Understanding how correlations can be used for quantum communication
protocols is a central goal of quantum information science. While
many authors have linked global measures of correlations such as entanglement
or discord to the performance of specific protocols, in general the
latter may require only correlations between specific observables.
In this work, we first introduce a general measure of correlations
for two-qubit states based on the classical mutual information between
local observables. We then  discuss the role of  the symmetry in the state's
correlations distribution and accordingly provide a classification
of maximally mixed marginals states (MMMS). We discuss the complementarity relation
between correlations and coherence. By focusing on a simple yet paradigmatic
example, i.e., the remote state preparation protocol, we introduce
a method to systematically define proper protocol-tailored measures
of correlations. The method is based on the identification of those
correlations that are relevant (useful) for the protocol. The approach
allows on one hand to discuss the role of the symmetry of the correlations
distribution in determining the efficiency of the protocol, both for
MMMS and general two-qubit quantum states, and on the other hand to
devise an optimized protocol for non-MMMS that can have a better efficiency
with respect to the standard one. The scheme we propose can be extended
to other communication protocols and more general bipartite settings.
Overall our findings clarify how the key resources in simple communication protocols
are the purity of the state used and the symmetry of correlations distribution.
\end{abstract}


\maketitle

\section{Introduction\label{sec:Introduction}}

The study of correlations in quantum systems has indeed a long, deep
and complex history. In particular, enormous efforts have been devoted
to characterizing the ``quantumness'' of correlations, or devising
measures of correlations aimed at capturing the ``quantum content''
of correlations present in a generic quantum state, such as quantum
entanglement\cite{HorodeckiReviewEntanglement} and quantum discord\cite{ModiReviewDiscord}.
Three premises underlie the derivation of such measures: $i)$ in
a quantum state there can be ``classical'' and ``quantum'' correlations
that coexist; $ii)$ it is possible to algorithmically identify and
separate the quantum vs the classical part of the correlations $iii)$
both parts can be quantified by means of a single number. In agreement
with these assumptions, the measures of correlations have been used
to establish a classification of\emph{ }quantum states\emph{ }based
on clear-cut distinction between quantum vs classical states (e.g.,
separable vs entangled states, discordant vs zero-discord states).
Furthermore, the correlation measures have been put in direct connection
with the efficiency of specific quantum protocols, as measured by
suitable figures-of-merit. An additional premise is implicit in this
effort: $iv)$ quantum correlations, interpreted as properties of
a given quantum state as a whole, underlie the efficiency of quantum
protocols. However, the strategy that follows the above premises is
sometimes unable to unequivocally provide a connection between the
performance of the protocol and a given measure of quantum correlations.
Therefore, the search for other perspectives is indeed possible and
it is in order. In particular, we propose to ``forget'' about the
quantum vs classical distinction, and rather focus on \emph{(classical)
correlations between sets of local observables}. Our proposal is based
on an idea that has been highlighted within the framework of the consistent
(decoherent) histories approach to quantum mechanics\cite{GriffithsConsistentHistoriesBook,OmnesConsistentHistories}
(and sometimes also within the standard interpretation\cite{PeresQuantumTheory}).
The state $\rho$ of a system, rather than a ``property'' of the
system, can be intended as a ``pre-probability'' i.e., a mathematical
device useful in order to calculate the probabilities of measurement
outcomes pertaining to (possibly incompatible) experiments. In a bipartite
setting for example, where A and B share a given state $ $$\rho$
and they want to implement a communication task, the probability distributions
pertaining to all pairs of local observables define the set of ``available
correlations'' stored in the state. When a specific protocol is enacted,
one is led to identify the subset of pairs of local observables that
are relevant for its realization, and therefore the corresponding
subset of \textit{relevant correlations}. In this sense, a bipartite
state can be imagined as a Multiple-Inputs/Multiple-Outputs system\cite{MIMO}
i.e., a communication system that can exploit several parallel channels
linking the transmitter and the receiver; the ``relevant channels''
are those identified by the pairs of local observables that are relevant
for a given protocol. In this perspective, on one hand\emph{ }quantum
states can be characterized as a whole by the average amount of (classical)
correlations between all pairs of local observables, whose value depends on the state purity,
and the symmetry
of the correlations distribution. On the other hand, the efficiency of specific\emph{
}quantum protocols\emph{ }can be connected with specific sets of local
observables and their mutual correlations. In this way one is able
to find protocol-specific measures of correlations and, as we demonstrate
in a specific example, to modify existing protocols in order to enhance
their efficiency. \\

While our approach is general and in principle applicable to multipartite
settings, in order to thoroughly examine the proposed strategy, here
we focus on the simplest case of quantum communication bipartite channels
provided by two-qubit quantum states $\rho_{AB}$, where the tensor
product structure $\mathcal{H=\mathbb{C}}^{2}\otimes\mathcal{\mathbb{C}}^{2}$
naturally provides the sets of local observables to study. In particular,
we start our analysis by focusing on states with maximally mixed marginals
(MMMS). The latter are particularly simple to study and yet they have
been widely used in the literature as prototypical instances of bipartite
communication channels \cite{HorodeckiReviewEntanglement,HorodeckiTetrahedron,ModiReviewDiscord}.
We will consider pairs of local von Neumann observables and their
correlations, as measured by the classical mutual information $\mathcal{I}$
of measurement outcomes. On the basis of $\mathcal{I}$, in the first
place we define a measure of the ``available correlations''by taking
a suitable average $\langle\mathcal{I}\rangle_{\Omega}$ over the
manifold $\Omega$ of local observables (which, in the case of two
qubits, are given by the product of two spheres $\Omega=S_{2}\times S_{2}$).
However, two states, with possibly different purities, can well have the same amount of average correlations
$\langle\mathcal{I}\rangle_{\Omega}$ (just as two states can have
the same amount of entanglement or discord) but \textit{they can be
strikingly different from the point of view of how the correlations
are distributed among the various observables}. In this perspective,
bipartite quantum states\emph{ }can be classified on the basis of
both the purity dependent quantity given by the \emph{average correlations},  
and by the purity independent feature given by the \emph{symmetry} of the correlations
distribution. Furthermore, it possible to introduce a relation between
the correlations of the pairs of observables and the coherence of
the product bases they define. In this respect we show that at fixed purity \emph{correlations
and coherence can be in general identified as complementary resources}.
\ \\ To assess the role of correlations in a quantum protocol, $\langle\mathcal{I}\rangle_{\Omega}$
may not be the most significant quantity. When one analyzes a given
communication task, one should spot out the set of observables that
are relevant for its realization. This is for example possible when
there exists a figure of merit $\mathcal{F}$ for the protocol that
explicitly depends on a specific subset of observables, i.e., a set
$\Omega_{RO}\subseteq\Omega$ of\emph{ relevant observables} (RO).
If this is the case, then one can immediately derive a protocol-related
measure of correlations by taking the average $\langle\mathcal{I}\rangle_{\Omega_{RO}}$
on this subset only. From the conceptual point of view, our perspective
is radically different from others: instead of considering an overall
property of the state, such as the entanglement, the discord or the
average mutual information $\langle\mathcal{I}\rangle_{\Omega}$,
we establish a direct connection between the (average) performance
of the protocol and the correlations pertaining to the relevant observables.
In the following we will fully develop a first example of this method
by applying it to the (two-qubit) remote state preparation (RSP) protocol
\cite{PatiRSP,BennettRSP,BennettRSP2,YeRSP}. The latter has been largely
studied in the literature and there have been many attempts to link
its performance to specific kinds of quantum correlations - such as
quantum discord \cite{DakicRSP} or entanglement \cite{horodeckiRSP}.
However, it has been showed that on one hand discord is neither sufficient
nor necessary for the efficiency of the protocol \cite{GiorgiRSP},
and on the other hand that states with lower content of entanglement
or discord can provide better efficiency than states with higher values
of both quantities \cite{ChavesRSP}. In our case, we will analyze
the protocol for both MMMS and general non-MMMS states. We will define
a functional $\mathcal{F}$ for RSP that allows us to identify the
set of relevant observables. While for states with maximally mixed
marginals (MMMS) all relevant observables are useful, i.e., they can
always be used to enhance $\mathcal{F}$, for general non-MMMS only
a subset of the relevant observables has this property. One can therefore
define the set of \textit{useful observables}\textit{\emph{ $\Omega_{U}\subseteq\Omega_{RO}$
and }}correspondingly introduce an alternative way of enacting RSP
based on useful observables only, such that the overall efficiency
of the protocol is improved. I\textit{\emph{n both cases (MMMS and
non-MMMS), }}we measure the advantage of using the correlations vs
not using them by means of a gain function $\mathcal{G}$ that explicitly
depends on the correlation of the useful observables. The average
gain will provide the link with the desired measure of correlations
pertaining to the protocol. \\
Throughout the whole discussion we analyze
how purity vs the symmetry of correlations affect the protocol. 
In general purity and symmetry of correlations can be thought as 
two fundamental resources: the purity fixes the amount of available correlations;
the symmetry determines how the correlations are distributed among the relevant observables.
As for symmetry alone, we finally show how it can be recognized as the key resource that allows
to establish the communication channel between the parties A and B
\emph{before }the state one wants to transfer is known. \\The paper
is organized as follows. In Section (\ref{sec:Preliminaries}) we
briefly define the formalism and the conventions used. In Section
(\ref{sec:Symmetry-and-(average) mutal info}) we introduce our measure
of correlations and we study the general properties of $\left\langle \mathcal{I}\right\rangle _{\Omega}$
and their relations with the state's symmetry for MMMS. Readers mainly
interested in RSP can skip this section and go directly to Section
(\ref{sec:Relevant-observables, pay off and gain}), where we discuss
in detail the RSP protocol for MMMS and non-MMMS. In Section (\ref{sec: RSP and symmetry})
we finally discuss the relation between symmetry and how freedom in
implementing the different steps of RSP. In Section (\ref{sec:Conclusions})
we derive our conclusions.

\section{Classification of quantum states based on correlations between observables}

We start by discussing how two-qubit quantum states can be characterized
on the basis of the pairwise correlations between local observables,
$\mathcal{I}(\hat{n},\hat{m})$. For simplicity, we focus on a subset
of states, those with maximally mixed marginals (MMMS). We show that
MMMS can be characterized by the average $\left\langle \mathcal{I}\right\rangle _{\Omega}$
as well as the symmetry of $\mathcal{I}(\hat{n},\hat{m})$, as defined
below. Finally, we discuss how the correlation content described by
$\mathcal{I}(\hat{n},\hat{m})$ is complementary to the coherence
of product basis defined by $\hat{n},\hat{m}$ in a given the state.

\subsection{Notation\label{sec:Preliminaries}}

By using the Bloch-Fano representation, one can show that an arbitrary
two-qubit state is equivalent, up to local unitary operations $U_{A}\otimes U_{B}$,
to the state:

\begin{equation}
\rho_{AB}=\frac{1}{4}(\mathbb{I}^{A}\otimes\mathbb{I}^{B}+\vec{a}\cdot\vec{\sigma}^{A}\otimes\mathbb{I}^{B}+\mathbb{I}^{A}\otimes\vec{b}\cdot\vec{\sigma}^{B}+\kappa\sum_{i}c_{i}\sigma_{i}^{A}\otimes\sigma_{i}^{B})
\end{equation}
where $\vec{a}\mathbf{=}|a|\hat{a}$ and $\vec{b}=|b|\hat{b}$ are
the Bloch vectors of the marginal states, and $E=\kappa\ \mbox{diag}(c_{1},c_{2},c_{3})$
is the correlation matrix in its diagonal form, and $\vec{\sigma}=\left(\sigma_{x},\sigma_{y},\sigma_{z}\right)^{T}$
is the vector of Pauli matrices. Therefore, the state is identified
by three vectors: the vectors $\vec{a},\vec{b}$ describing the reduced
density matrices $\rho_{A},\rho_{B}$ and the correlation vector $\vec{c}=\kappa\hat{c}=\kappa(c_{1},c_{2},c_{3}),\ \kappa=|\vec{c}|$.
In the following, we will focus on maximally-mixed marginal states
(MMMS), defined as the states with for which $\vec{a}=\vec{b}=\vec{0}$,
and which hence have maximally mixed reduced states on $\rho_{A}=\rho_{B}=\frac{1}{2}\mathbb{I}$:

\[
\rho_{AB}^{(MMMS)}=\frac{1}{4}(\mathbb{I}^{A}\otimes\mathbb{I}^{B}+\kappa\sum_{i}c_{i}\sigma_{i}^{A}\otimes\sigma_{i}^{B}).
\]
MMMS are completely characterized by the correlation vector $\vec{c}$.
The condition for $\rho_{AB}$ to be a good quantum state is the positivity
condition $\rho_{AB}>0$ . The latter implies that $\vec{c}\in\mathcal{T}$
i.e., $\vec{c}$ is a vector in $\mathbb{R}^{3}$ contained in the
tetrahedron $\mathcal{T}$ with vertices $(-1,-1,-1),(-1,1,1),(1,-1,1),(1,1,-1)$\cite{HorodeckiTetrahedron}.
The value of the parameter $\kappa$ defines the purity of the state
that reads $(1+\kappa^{2})/4$.\ \\
In the following we will focus on pairs of von Neumann observables.
The latter are operators that can be represented as $O_{A(B)}=\sum_{k}o_{k}\Pi_{k}^{A(B)}$
, $\left\{ \Pi_{k}^{A(B)}\right\} $ being a complete orthogonal set
of projectors on the Hibert space $\mathcal{H}_{A(B)}$. Since we
are dealing with qubits any projector can be written in terms of Pauli
matrices as 
\[
\Pi_{\pm}^{A(B)}(\hat{m})=\left(\mathbb{I}\pm\hat{m}\cdot\vec{\sigma}\right)/2
\]
where $\hat{m}$ is a unit vector belonging to a single qubit Bloch
sphere, $\vec{\sigma}=\left(\sigma_{x},\sigma_{y},\sigma_{z}\right)$
and $\Pi_{\pm}=\ket{\pm\hat{m}}\bra{\pm\hat{m}}$. We are interested
in the correlations between pairs of observables $\hat{n}\cdot\vec{\sigma},\hat{m}\cdot\vec{\sigma}$
(pertaining to the subsystem $A$ and $B$ respectively), whose projectors
are defined as $\Pi_{\pm}^{A}(\hat{n}),\Pi_{\pm}^{B}(\hat{m})$ .
The measure of correlations we use is the standard classical mutual
information $\mathcal{I}(\Pi_{\pm}^{A}(\hat{n}),\Pi_{\pm}^{B}(\hat{m}))\equiv\mathcal{I}(\hat{n},\hat{m})$,
which can be written in terms of the joint probability distribution
\[
p_{ij}=Tr\left[\rho_{AB}\Pi_{i}^{A}(\hat{n})\otimes\Pi_{j}^{B}(\hat{m})\right],\qquad i,j=\pm
\]
and of the marginals $p_{i}=Tr\left[\rho_{A}\Pi_{i}^{A}(\hat{n})\right]$,
$p_{j}=Tr\left[\rho_{B}\Pi_{j}^{B}(\hat{m})\right]$ as 
\[
\mathcal{I}(\hat{n},\hat{m})=-\sum_{i}p_{i}\log_{2}p_{i}-\sum_{j}p_{j}\log_{2}p_{j}+\sum_{ij}p_{ij}\log_{2}p_{ij}
\]

For MMMS the probability for the joint measurements defined by $\left(\hat{n},\hat{m}\right)$
can be expressed in terms of the correlations matrix as $p_{i,j}=\left(1+ij\ \hat{m}E\hat{n}^{T}\right)/4$,
$i,j=\pm$, whereas the probabilities for the single local measurements
yield $p_{\pm}^{A(B)}=1/2$.

\subsection{Symmetry and distribution of correlations\label{sec:Symmetry-and-(average) mutal info}}

With the above notations, the mutual information between two local
observables $\hat{n},\hat{m}$ in MMMS can be simply expressed as

\[
\mathcal{I}(\hat{n},\hat{m})=\frac{1}{2}(1-x)\log_{2}(1-x)+(1+x)\log_{2}(1+x))
\]
where $x=\hat{n}E\hat{m}^{T}=\kappa\ \hat{n}\mbox{diag}(c_{1},c_{2},c_{3})\hat{m}^{T}$.
From this formula, it immediately follows that the correlations between
any two observables are a monotonic function of $\kappa$ i.e., of the purity, and that
for any fixed $\kappa$ the distribution of correlations between different
pairs of observables depends on the direction of the correlation vector
$\hat{c}$. States, identified by their $\hat{c}$, can be classified
on the basis of the distribution of correlations they yield. \\
 \ \\
A first classification of the states and the corresponding directions
$\hat{c}$ can be done on the basis of the local symmetries of the
correlations, that follow from the local symmetries for the state.
A state $\rho$ has a local unitary symmetry if there are local unitaries
$U_{A}\otimes U_{B}$ such that $U_{A}\otimes U_{B}\rho U_{A}^{\dagger}\otimes U_{B}^{\dagger}=\rho$.
The local unitary symmetries of the state form a group $\mathcal{LU}$
called Local Unitary Stabilizer \cite{Lyons}, which is a discrete
or continuous subgroup of $SU(2)\otimes SU(2)$. Local unitaries $U\in SU(2)$
acting on the Hilbert space can be mapped to rotations $O\in SO(3)$
acting on the Bloch sphere: indeed, there exists a (unique) rotation
$O\in SO(3)$ such that $U\hat{n}\cdot\vec{\sigma}U^{\dagger}=(O\hat{n})\cdot\vec{\sigma}$.
By virtue of this $SU(2)\to SO(3)$ mapping, local unitary symmetries
can be expressed in terms of \textit{special orthogonal} transformations
that leave the correlation matrix invariant: 
\begin{equation}
O_{A}EO_{B}^{T}=E\label{eq: local invariance so(3)}
\end{equation}
where $O_{A},O_{B}\in SO(3)$. The fact that a state defined by $\vec{c}$
has symmetry group $\mathcal{LU}$ can be viewed in two equivalent
ways. On one hand, for all $\hat{n},\hat{m}$ also $\mathcal{I}(\hat{n},\hat{m})(\vec{c})$
is left invariant by the action of $\mathcal{LU}$ on $\rho$. On
the other hand, local symmetries of the state imply a symmetry in
the distribution of correlations: given a pair of local observables
$(\hat{n},\hat{m})$, all the pairs $(\hat{n}',\hat{m}')=(\hat{n}O_{A},\hat{m}O_{B})$
have the same value of mutual information. \\
 Given a direction $\hat{c}$ with a specific $\mathcal{LU}$, we
are interested in identifying the equivalence class of directions
that for fixed $\kappa$ (purity) yield isomorphic distributions of correlations.
Formally, for any fixed $\kappa$ and any given $\hat{c}$, we want
to identify the directions $\hat{d}$ such that for any pair of observables
$(\hat{n},\hat{m})$ there exists a pair of observables $(\hat{n}',\hat{m}')$
such that $\mathcal{I}(\hat{n},\hat{m})(\kappa\hat{c})=\mathcal{I}(\hat{n}',\hat{m}')(\kappa\hat{d})$,
i.e., there exists a bijective map $\phi:(\hat{n},\hat{m})\rightarrow(\hat{n}',\hat{m}')$,
realizing a change of local coordinates on the Bloch spheres, such
that $\mathcal{I}(\hat{n},\hat{m})(\kappa\hat{c})=\mathcal{I}(\phi(\hat{n},\hat{m}))(\kappa\hat{d})$.
Thus, given a direction $\hat{c}=(c_{1},c_{2},c_{3})$, we want to
identify the following equivalence class $\mathcal{LU}_{\hat{c}}^{eq}$
of directions $\hat{d}=(d_{1},d_{2},d_{3})$:

\begin{align}
\mathcal{LU}_{\hat{c}}^{eq}\equiv\big\{\hat{d}\ :\ \exists\kappa\ |\ \forall(\hat{m},\hat{n}),\ \exists(\hat{m}',\hat{n}')\ |\ \label{eq: LUeq0}\\
\mathcal{I}(\hat{n},\hat{m})(\kappa\hat{d})=\mathcal{I}(\hat{n}',\hat{m}')(\kappa\hat{c})\big\}\nonumber 
\end{align}
In order to identify the components of a given class one has to notice
that a local change of coordinates on the Bloch spheres $S_{2}\times S_{2}$
corresponds to a pair of now \textit{orthogonal} transformations $O_{A},O_{B}\in O(3)$
acting on $\hat{n},\hat{m}$ as $\hat{n}'=\hat{n}O_{A}$ and $\hat{m}'=\hat{m}O_{B}$.
In order to have $\mathcal{I}(\hat{n},\hat{m})(\kappa\hat{d})=\mathcal{I}(\hat{n}',\hat{m}')(\kappa\hat{c})$
we must have 
\[
|\hat{n}'\mbox{diag}(d_{1},d_{2},d_{3})\hat{m}'^{T}|=|\hat{n}\mbox{diag}(c_{1},c_{2},c_{3})\hat{m}^{T}|
\]
which can be rewritten as 
\begin{equation}
O_{A}\mbox{diag}(c_{1},c_{2},c_{3})O_{B}^{T}=\pm\mbox{diag}(d_{1},d_{2},d_{3})\label{eq:samedistribution}
\end{equation}
Equation (\ref{eq:samedistribution}) severely constrains the form
of $\hat{d}$. Indeed, since the matrices $\mbox{diag}(c_{1},c_{2},c_{3})$
and $\mbox{diag}(d_{1},d_{2},d_{3})$ are related by two orthogonal
rotations as above, they must have the same singular values. This
implies that $|d_{1}|,|d_{2}|,|d_{3}|$ are related to $|c_{1}|,|c_{2}|,|c_{3}|$
by a permutation. As a result, we must have 
\begin{align}
\mathcal{LU}_{\hat{c}}^{eq}=\big\{\hat{d}=(s_{1}c_{\sigma(1)},s_{2}c_{\sigma(2)},s_{3}c_{\sigma(3)})|\label{eq: LUeq}\\
s_{i}\in\{-1,1\},\vec{\sigma}\in P(1,2,3)\big\}\nonumber 
\end{align}
where $P(1,2,3)$ is the set of permutations of three indices. $\mathcal{LU}_{\hat{c}}^{eq}$
can be seen as the orbit of a discrete subgroup of $O(3)$ that acts
on the given $\hat{c}$ and is isomorphic to $G\sim S_{3}\otimes E_{8}$,
where $S_{3}$ is the symmetric group of order $3$, corresponding
to the permutations of three indices, and $E_{8}$ is the elementary
Abelian group of order $8$ that realizes the changes of signs $s_{i}$
in Eq. (\ref{eq: LUeq}). As discussed in the Appendix \ref{sec:Appendix-A},
the transformations in $G$ can be realized by a combination of local
unitary rotations and a non-unitary local spin flip that implements
the transformation $\hat{c}\rightarrow-\hat{c}$; furthermore, the
total number of \textit{different} equivalent directions $|\mathcal{LU}_{\hat{c}}^{eq}|\le48$
depends on the specific $\mathcal{LU}$ and $\hat{c}$. \\
In Ref.\cite{Lyons} a complete classification of the continuous $\mathcal{LU}$
for $N$-qubit states was given; starting from such classification
we identify the following classes of MMMS: 
\begin{enumerate}
\item $\rho_{3iso}$ states (``isotropic states''): they belong the class
$\mathcal{LU}_{\hat{c}}^{eq}$ with $\hat{c}=(1,1,1)/\sqrt{3}$. These
states are invariant with respect to local unitaries of the kind $U\otimes U,\ U\in SU(2)$
and we define the class as $\mathcal{LU}_{3iso}$; it holds $|\mathcal{LU}_{3iso}|=8$.
Bell and Werner states belong to this class of isotropic states. 
\item $\rho_{2iso}^{\epsilon}$ states: they are equivalent to $\hat{c}=(\epsilon,\epsilon,\sqrt{1-2\epsilon^{2}}),\ 0<\epsilon^{2}\le1/2$;
these states are invariant with respect to that subset of local unitaries
of the kind $U\otimes U,\ U=\exp-i\theta\sigma_{z}^{A(B)}$ ; we define
the class as $\mathcal{LU}_{2iso}^{\epsilon}$, which has $|\mathcal{LU}_{2iso}^{\epsilon}|=24$
elements if $\epsilon\neq0$ and $12$ elements if $\epsilon^{2}=1/2$. 
\item $\rho_{2iso}^{0}$ states: they are equivalent to $\hat{c}=(0,0,1)$;
these states are invariant with respect to that subset of local unitaries
of the kind $U_{A}\otimes U_{B},\ U_{A}=\exp-i\theta\sigma_{z}^{A},\ U_{B}=\exp-i\eta\sigma_{z}^{B}$
where in general $\theta\neq\eta$. we define the class as $\mathcal{LU}_{2iso}^{0}$,
which has $|\mathcal{LU}_{2iso}^{0}|=6$. This class coincides with
the MMMS states that are called ``classical'' in the literature
because they are diagonal in a product basis and have zero quantum
discord. 
\end{enumerate}
The above classes constitute a fine-graining of the Local Stabilizer
formalism. For example, while in our case $\mathcal{LU}_{2iso}^{\epsilon}$
define different classes for different values of $\epsilon$, since
they give rise to inequivalent distribution of correlations, they
are all equivalent in the Local stabilizer setting.

\subsection{Average correlations}

\begin{figure}
\includegraphics[scale=0.5]{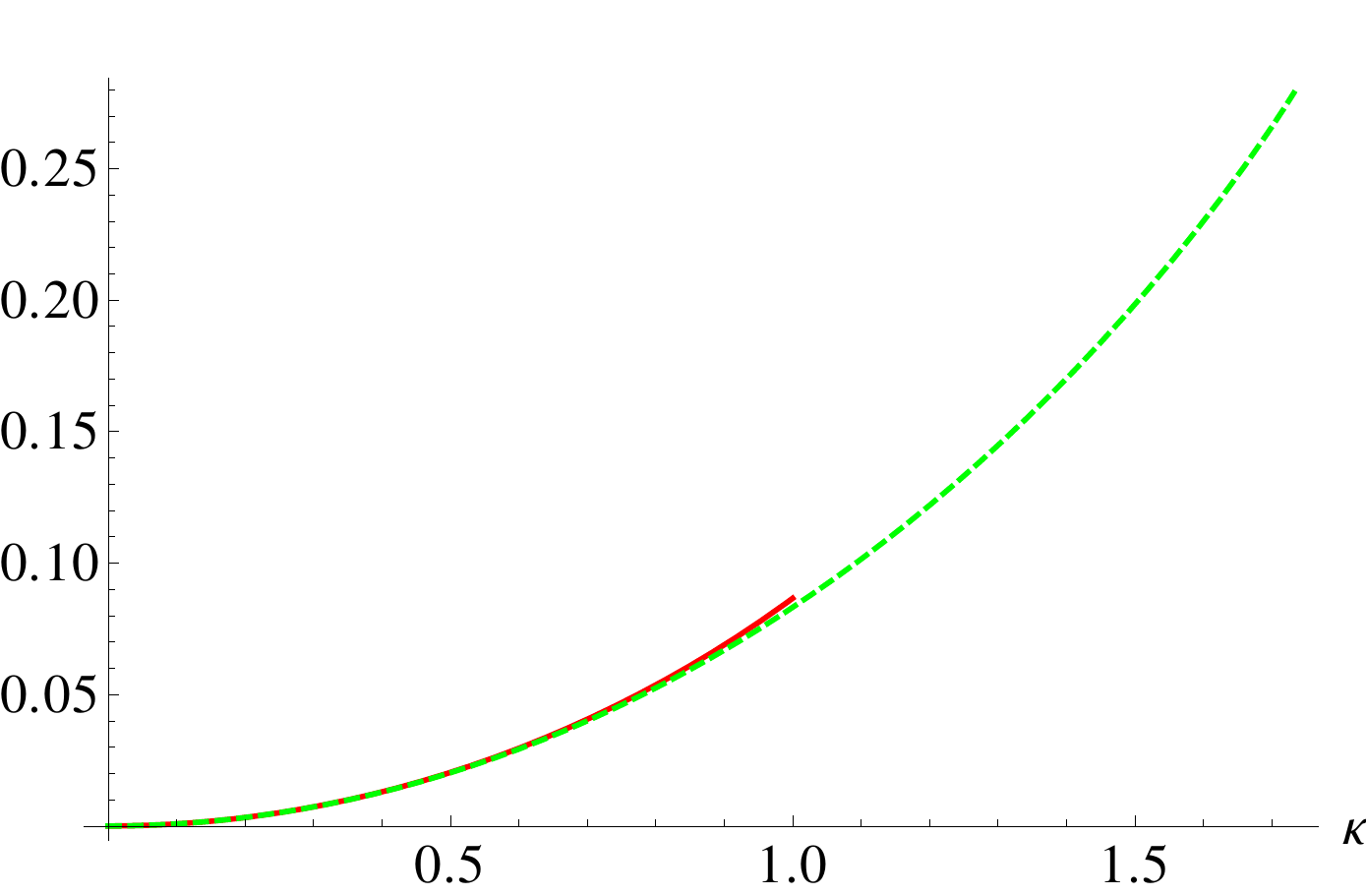}\protect\caption{Average mutual information $\left\langle \mathcal{I}\right\rangle _{\Omega}(\kappa)$
for $\rho_{2iso}^{0}$ (red) and $\rho_{3iso}$ (green). The average
shows a weak dependence on the symmetry class of the states (see text
for discussion)}
\end{figure}

Given the above classification we now pass to analyze the average
amount of pairwise correlations between observables as measured by
the average mutual information

\begin{equation}
\left\langle \mathcal{I}\right\rangle _{\Omega}=\frac{1}{2}\left\langle (1-x)\log_{2}(1-x)+(1+x)\log_{2}(1+x))\right\rangle _{\Omega}
\end{equation}
where the average \cite{TranDakic} is taken over $\hat{m},\hat{n}\in\Omega=S^{2}\times S^{2}$
i.e., the Bloch spheres for the two qubits where the observables are
identified by the unit vectors $\hat{n},\hat{m}$. The study of this
function will allow us to identify, among the above classes of states,
those that are extremal with respect $\left\langle \mathcal{I}\right\rangle _{\Omega}$.
Evidently, for a fixed direction of the correlation vector $\hat{c}$
the average $\left\langle \mathcal{I}\right\rangle _{\Omega}$ is
a growing function of $\kappa$ i.e., of the state purity. In order to perform the average,
we can first evaluate the average over $\hat{n}$ only. To this aim,
we use the expansion $\ln(1+x)=\sum_{n=1}^{\infty}\frac{(-1)^{(n+1)}}{n}x^{n}$
and the fact that $\left\langle x^{2n+1}\right\rangle _{(\hat{n})}=0$
to obtain 
\begin{align*}
 & \frac{1}{2}\left\langle (1+x)\log_{2}(1+x)+(1-x)\log_{2}(1-x)\right\rangle _{\hat{n}}=\\
= & \mbox{\ensuremath{\frac{1}{(2\ln2)}}}\sum_{h=1}^{\infty}\frac{\langle x^{2h}\rangle_{\hat{n}}}{h(2h-1)}
\end{align*}

\begin{figure}
\includegraphics[scale=0.6]{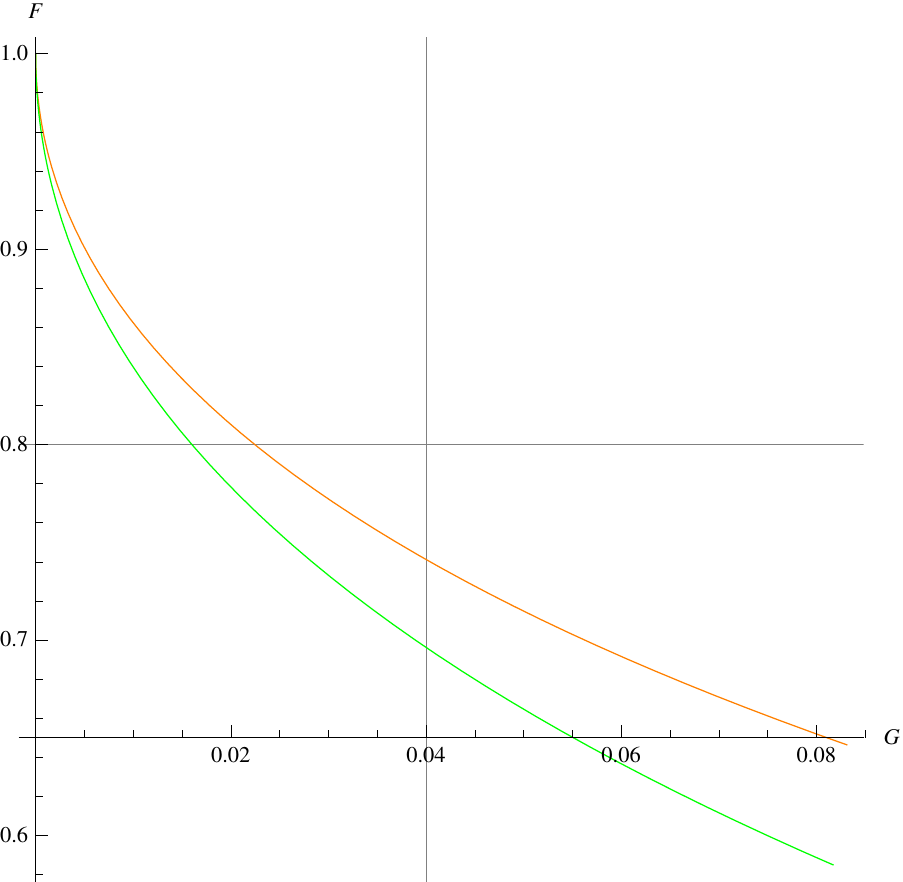}
\protect\caption{figure-of-merit $\left\langle \mathcal{F}\right\rangle _{\Omega_{RO}}$
vs gain $\left\langle \mathcal{G}\right\rangle _{\Omega_{RO}}$ for
$\rho_{2iso}^{0}$ (orange) and $\rho_{3iso}$ (green) and $0<\kappa<1$.
At fixed $\left\langle \mathcal{G}\right\rangle _{\Omega_{RO}}$ (vertical
grey line), $\left\langle \mathcal{F}\right\rangle _{\Omega_{RO}}$
is lower for $\rho_{3iso}$ . At fixed $\left\langle \mathcal{F}\right\rangle _{\Omega_{RO}}$(horizontal
grey line), $\left\langle \mathcal{G}\right\rangle _{\Omega_{RO}}$
is lower for $\rho_{3iso}$ }
\label{Fig. 2}
\end{figure}

Integrating with respect to $\hat{n}$, we get $\left\langle x^{2h}\right\rangle _{\hat{n}}=\frac{1}{2h+1}R^{h}$,
with $R=(c_{1}^{2}m_{1}^{2}+c_{2}^{2}m_{2}^{2}+c_{3}^{2}m_{3}^{2})$.
Upon resumming the series, the overall average mutual information
$\left\langle \mathcal{I}\right\rangle _{S^{2}}(\hat{m})$ for a single
observable $\hat{m}$ can be thus evaluated as

\begin{equation}
\left\langle \mathcal{I}\right\rangle _{S^{2}}(\hat{m})=\frac{(1+R)\:\text{atanh}\sqrt{R}-\sqrt{R}(1-\ln(1-R))}{\sqrt{R}\ln4}\label{eq: Average Mutual Info Single Observable}
\end{equation}
$\left\langle \mathcal{I}\right\rangle _{S^{2}}(\hat{m})$ is a monotonically
growing function of $R=\hat{m}EE^{T}\hat{m}^{T}$ and we have $0\leq\left\langle \mathcal{I}\right\rangle _{\hat{n}}(\hat{m})\leq0.27865$
. The average mutual information $\left\langle \mathcal{I}\right\rangle _{\Omega}$
can be obtained by further averaging with respect to $\hat{m}$. The
average can be expressed analytically only in simple cases. For states
$\rho_{2iso}^{0}$ , we get

\begin{eqnarray}
\left\langle \mathcal{I}\right\rangle _{\Omega}(\kappa)=\frac{-6\kappa+\left(6+2\kappa^{2}\right)\text{atanh}\kappa}{8\kappa\log2}+\label{eq: Average Mutual info single c_i}\\
\frac{\kappa^{3}\text{\ensuremath{\Phi}}(\kappa^{2},2,\frac{3}{2})+4\kappa\log(1-\kappa^{2})}{8\kappa\log2}\nonumber 
\end{eqnarray}
where $\Phi$ is the Lerch transcendent function; for the isotropic
states $\rho_{3iso}$,

\begin{eqnarray}
\left\langle \mathcal{I}\right\rangle _{\Omega}(\kappa)=\frac{(3+\kappa^{2})\text{atanh}(\kappa/\sqrt{3})}{\sqrt{3}\kappa\log4}\label{eq: Average Mutual info Isotropic states}\\
-\frac{\sqrt{3}\kappa\left(1-\log\left(1-\kappa^{2}/3\right)\right)}{\sqrt{3}\kappa\log4}\nonumber 
\end{eqnarray}
At fixed $\kappa$, $\rho_{2iso}^{0}$ and $\rho_{3iso}$ are found
to be extremal in terms of the average correlations. Indeed, one can
study some general properties of MMMs with respect to $\left\langle \mathcal{I}\right\rangle _{\Omega}(\kappa\hat{c})$
as a function of $\hat{c}=(\sin\alpha\cos\beta,\sin\alpha\sin\beta,\cos\beta)$.
The results can be summarized in the following proposition.
\begin{prop}
For fixed $\kappa\le1$ he states with minimal $\left\langle \mathcal{I}\right\rangle _{\Omega}(\kappa\hat{c})$
are $\rho_{3iso}$ and the states with maximal $\left\langle \mathcal{I}\right\rangle _{\Omega}(\kappa\hat{c})$
are $\rho_{2iso}^{0}$. If $\kappa\ge1$ , the minima remain in correspondence
of $\rho_{3iso}$, while the maxima are to be found on the intersection
between the sphere of radius $\kappa$ and the tetrahedron $\mathcal{T}$. 
\end{prop}
The proof of Proposition 1 can be found in Appendix \ref{sec:Appendix-B}.
In Fig. 1 we plot $\left\langle \mathcal{I}\right\rangle _{\Omega}(\kappa\hat{c})$
for $\rho_{3iso}$ and $\rho_{2iso}^{0}$. From this plot, one can
see that for $\kappa\le1$ the value of $\left\langle \mathcal{I}\right\rangle _{\Omega}(\kappa\hat{c})$
is essentially determined by the purity of the state and has weak dependence on
the direction $\hat{c}$. This fact dims the relevance of the symmetry
properties of the correlations distribution, which becomes quite evident
when one considers a specific communication protocol, for which only
a specific subset of correlations is relevant. For example, the effect
of the symmetry is very apparent when one considers the subset of
maximally correlated observables i.e., the subset $\Omega_{Max}\subset S_{2}\times S_{2}$
defined by $\Omega_{Max}=\left\{ \left(\hat{n}_{M},\hat{m}_{M}\right)|\hat{n}_{M}E\hat{m}_{M}=\max_{\hat{n},\hat{m}}\hat{n}E\hat{m}\right\} $.
For the classes $\mathcal{LU}^{eq}$ identified above: 
\begin{itemize}
\item for the $\rho_{2iso}^{0}$\emph{, $\Omega_{Max}$} is defined by the
equation $n_{3}m_{3}=1$ which is satisfied only if $\hat{n}=(0,0,\pm1),\hat{m}=(0,0,\pm1)$.
We have $\Omega_{Max}=\{\hat{n}=(0,0,\pm1),\hat{m}=(0,0,\pm1)\}$
with $\dim\Omega_{Max}=0$. 
\item for the $\rho_{2iso}^{\epsilon}$ and $\epsilon\in\left(1/\sqrt{3},1/\sqrt{2}\right)$,
\emph{$\Omega_{Max}$} is defined by the equation $n_{1}m_{1}+n_{2}m_{2}=1$
, which is satisfied if the directions of both observables lie on
the equatorial circle $S_{1}$ (i.e.,$m_{3}=n_{3}=0$) and are coincident.
We have $\Omega_{Max}\sim S_{1}$ with $\dim\Omega_{Max}=1$. 
\item for the \emph{$\rho_{3iso}$, $\Omega_{Max}$} is defined by the equation
$\hat{n}\cdot\hat{m}=1$, which is satisfied if the direction of the
two observables coincide. Thus, $\Omega_{Max}\sim S_{2}$ with $\dim\Omega_{Max}=2$. 
\end{itemize}
It is evident that therefore the symmetries can have important implications
for protocols based on maximally correlated observables,as it will
become clear in the discussion about RSP, see for example Figures
(\ref{Fig. 2}) and (\ref{Fig. 3}) and the related discussion.

\subsection{Complementarity between correlations and coherence}

An important aspect of the correlations between between observables
$(\hat{n},\hat{m})$ is that they can be seen as complementary to
the coherence properties of the product basis identified by $\hat{n},\hat{m}$
i.e., $\mathcal{B}_{(\hat{n},\hat{m})}=\left\{ \ket{\pm\hat{n}}\ket{\pm\hat{m}}\right\} $,
with respect to the given state. In order to assess this point one
can use the coherence function\cite{BaumgrazCoherence,OurCoherence}
given by $Coh_{\mathcal{B}_{(\hat{n},\hat{m})}}(\vec{c})=\mathcal{H}_{\mathcal{B}_{(\hat{n},\hat{m})}}(\vec{c})-\mathcal{S}(\rho)$
where $\mathcal{H}_{\mathcal{B}_{(\hat{n},\hat{m})}}(\vec{c})$ is
the entropy of the joint probability distribution obtained by a measurement
of the observables identified by $\hat{n},\hat{m}$. For MMMS, we
obtain 
\begin{equation}
Coh_{\mathcal{B}_{(\hat{n},\hat{m})}}(\vec{c})=2-\mathcal{I}(\hat{n},\hat{m})(\vec{c})-\mathcal{S}(\rho)\label{eq: coherence vs MI}
\end{equation}
with $\mathcal{S}(\rho)$ the von Neumann entropy of $\rho$. The
above formula establishes a clear link between the correlations between
local observables and the coherence of the product bases they define.
Therefore, the coherence properties for MMMS can be inferred from$\mathcal{I}(\hat{n},\hat{m})$
and $\left\langle \mathcal{I}\right\rangle _{\Omega}$. We obtain
that 
\begin{prop}
i) for fixed $\hat{c}$, $Coh_{\mathcal{B}_{(\hat{n},\hat{m})}}(\vec{c})$
and $\left\langle Coh(\vec{c})\right\rangle $ are a growing function
of $\kappa$ i.e., of the purity of the state; ii) at fixed $\kappa$,
the higher the correlations between pairs of observables $(\hat{n},\hat{m})$
the lower their coherence with respect to the global state $\rho$;
iii) at fixed $\kappa$, for all states such that $\hat{c}\in\mathcal{LU}_{\hat{c}}^{eq}$,
$Coh_{\mathcal{B}_{(\hat{n},\hat{m})}}$ enjoys the $\mathcal{LU}$
symmetry; iv) at fixed $\kappa$, in general $Coh_{\mathcal{B}_{(\hat{n},\hat{m})}}(\vec{c})\neq Coh_{\mathcal{B}_{(\hat{n},\hat{m})}}(-\vec{c})$,
since in general $\mathcal{S}(\rho_{\vec{c}})\neq\mathcal{S}(\rho_{-\vec{c}})$,
and therefore each equivalence class splits as $\mathcal{LU}_{\hat{c}}^{eq}=\tilde{\mathcal{LU}}_{+\hat{c}}^{eq}\bigcup\tilde{\mathcal{LU}}_{-\hat{c}}^{eq}$
v) \textup{all the states such that $\hat{d}\in\tilde{\mathcal{LU}}_{+\hat{c}}^{eq}\left(\tilde{\mathcal{LU}}_{-\hat{c}}^{eq}\right)$
have the same value of }$\left\langle Coh(\vec{c})\right\rangle \left(\left\langle Coh(-\vec{c})\right\rangle \right)$.
\\

\end{prop}
The first property simply stems from the fact that the coherence function
$Coh_{\mathcal{B}_{(\hat{n},\hat{m})}}(\vec{c})=\mathcal{H}_{\mathcal{B}_{(\hat{n},\hat{m})}}-\mathcal{S}(\rho)$,
since $\mathcal{H}_{\mathcal{B}_{(\hat{n},\hat{m})}}$ is a growing
function of $\kappa$ and $\mathcal{S}(\rho)$ is a decreasing function
of $\kappa$. \\
The second property is quite relevant since it can be stated as: for
pairs of observables $(\hat{n},\hat{m})$ \textit{correlations and
coherence are complementary properties}. In particular, for pure (Bell)
states the pairs $(\hat{n},\hat{m})\in\Omega_{Max}$ that have maximal
mutual information have minimal coherence. Therefore communication
protocols involving MMMS and that are based on $(\hat{n},\hat{m})$
pairs can in principle be divided in two different categories: those
that rely on correlations and those that rely on coherence. Although
this subdivision is in principle sharp, we will see that the RSP protocol
for example falls in the first category. In~\cite{OurCoherence2} we have provided an example of protocol that falls in the second category: quantum phase estimation, which turns out to be based on coherence rather than correlations. 
\\
 The third property descends from the fact that $\mathcal{S}(\rho)$
is invariant with respect to any unitary rotation in $SU(4)$, and
it allows to extend the discussion already made about $\mathcal{I}_{\rho}(\hat{n},\hat{m})(\vec{c})$
and $\left\langle \mathcal{I}_{\rho}(\vec{c})\right\rangle $ to $Coh_{\mathcal{B}_{(\hat{n},\hat{m})}}(\vec{c})$
and $\left\langle Coh(\vec{c})\right\rangle $ (where the average
is taken over the two Bloch spheres) since they inherit the same symmetry
properties. \\
The fourth property marks a difference between the set of states that
are locally unitarily equivalent to $\vec{c}=\kappa\hat{c}$ and those
that are unitarily equivalent to $\vec{c}=-\kappa\hat{c}$: they both
have the same purity, and therefore same linear entropy, but in general
different $\mathcal{S}(\rho_{\vec{c}})$, since the transformation
$\hat{c}\rightarrow-\hat{c}$ does not preserve the spectrum of $\rho$.
For the states that have the higher $\mathcal{S}(\rho_{\vec{c}})$
the pairs $(\hat{n},\hat{m})$ have the lower coherence; a property
which is consistent with the fact that states with higher values of
$\mathcal{S}(\rho_{\vec{c}})$ are more ``mixed'' or entropic when
one considers them in terms of their global $SU(4)$ property $\mathcal{S}(\rho)$
that depends on the spectrum. \\
 The fifth property is analogous to the same property for $\mathcal{I}_{\rho}(\hat{n},\hat{m})(\vec{c})$
and $\left\langle \mathcal{I}_{\rho}(\vec{c})\right\rangle $, since
$\mathcal{S}(\rho_{\vec{c}})\left(\mathcal{S}(\rho_{-\vec{c}})\right)$
is constant for fixed $\kappa$.

\section{Relevant observables, useful correlations and performance in RSP\label{sec:Relevant-observables, pay off and gain}}

We are now ready to introduce the main quantifiers necessary for the
description of how the correlations are used in a the remote state
preparation protocol. We first define the figure-of-merit $\mathcal{F}$,
we optimize it and we find out what the \emph{relevant observables}
for the protocol are. This will allow us to introduce the \emph{gain}
$\mathcal{G}$ that measures the advantage in using the correlations
in the protocol. While we mainly focus our discussions to the relevant
classes of states previously defined, the tools and procedures we
outline can in general be applied to any two-qubit state.

\subsection{Remote state preparation}

Let us start from with a brief review of the remote state preparation
(RSP) protocol. Starting from a state $\varrho_{AB}$, two parties
$A$ and $B$ wish to prepare on $B$ side an arbitrary pure state
$\ket{\hat{n}}$ belonging to the Bloch sphere circle orthogonal to
a given Bloch sphere axis $\hat{\beta}$, where $\hat{n}$ is the
vector identifying the state in the Bloch sphere of $B$, such that
$\hat{n}\cdot\hat{\beta}=0$ (note that here and in the following
we will use $\hat{n}$ both for the state $\ket{\hat{n}}$ and the
observable $\hat{n}\cdot\vec{\sigma}$; the meaning will be clear
from the context). To prepare state $\hat{n}$ on $B$, $A$ performs
a local measurement on her qubit corresponding to the observable $\hat{m}\cdot\vec{\sigma}$.
Depending on the outcome $i=\pm1$, the conditional post measurement
states of $B$ are identified by the vectors 
\begin{equation}
\vec{r}_{i}=\frac{\vec{b}+i\hat{m}E^{T}}{2p_{i}^{A}}.\label{eq: RST postmeasurment state-1}
\end{equation}
where $p_{i}^{A}=\frac{1}{2}(1+i\hat{m}\cdot\vec{a)}$. Upon measuring,
$A$ sends a classical message to $B$ revealing the measurement outcome
$i$. If $i=1$, $B$ leaves his qubit unperturbed; if $i=-1$ he
performs a rotation of $\pi$ around the axis $\hat{\beta}$, $R^{\pi}(\hat{\beta})$
. Taking into account $B$'s conditional rotations the state in $B$
is:

\begin{equation}
\tilde{\varrho}^{B}(\hat{m})=p_{1}^{A}\varrho_{B|1}+p_{-1}^{A}R^{\pi}(\hat{\beta})\varrho_{B|-1}\label{eq:conditionalaveragestate-1}
\end{equation}
where $\varrho_{B|i}$ are the corresponding post measurement states
identified by $\vec{r}_{i}$. The state $\tilde{\varrho}^{B}(\hat{m})$
is identified by the Bloch vector 
\begin{equation}
\vec{r}=\hat{m}E^{T}+\big(\vec{b}-\hat{m}E^{T}\big)\cdot\hat{\beta}\ \hat{\beta}\label{eq: RSP post-measurement states}
\end{equation}
The effectiveness of the protocol depends on how close $\vec{r}$
is to the target state $\hat{n}$.

\subsection{Figure-of-merit, relevant observables and gain for MMMS\label{sub: MMMS pay-off, gain, relevant observables}}

We start to now analyze the RSP protocol for MMMS and later extend
the results to the other classes of states. For MMMS, we have 
\[
p_{i}^{A}=\frac{1}{2}\qquad\vec{r}_{i}=i\hat{m}E^{T}\qquad\vec{r}=\hat{m}E^{T}-\big(\hat{m}E^{T}\big)\cdot\hat{\beta}\ \hat{\beta}
\]
We first want estimate the efficiency of the RSP procedure. One natural
possibility is to compare the probabilities of a $\ket{\pm\hat{n}}$
measurement performed by \textbf{$B$} on: $i)$ the desired output
state $+\hat{n}$ i.e., $p_{+}=1,p_{-}=0$; $ii)$ the actual output
of the protocol $\vec{r}$ i.e., $p_{\pm}^{E}=\left(1\pm\hat{n}\cdot\vec{r}\right)/2=\left(1\pm\hat{n}E\hat{m}^{T}\right)/2$.
We therefore define as the relevant figure-of-merit the relative entropy
between these probability distributions:

\begin{equation}
\mathcal{\mathcal{F}}(\hat{n},\hat{m})=p_{+}\log_{2}\frac{p_{+}}{p_{+}^{E}}+p_{-}\log_{2}\frac{p_{-}}{p_{-}^{E}}=1-\log_{2}(1+\hat{n}E\hat{m}^{T})\label{eq: relative entropy as pay off}
\end{equation}
This function describes how much the probability distribution given
by a measurement of $\hat{n}$ onto $\vec{r}$ is statistically distinguishable
from the probability distribution given by $p_{+}=1,p_{-}=0$. One
has that $0\le\mathcal{\mathcal{F}}(\hat{n},\hat{m})$; $\mathcal{\mathcal{F}}(\hat{n},\hat{m})=0$
when $\vec{r}=\hat{n}$; $\mathcal{\mathcal{F}}(\hat{n},\hat{m})=1$
when $\hat{n}\cdot\vec{r}=0$; and $\mathcal{\mathcal{F}}(\hat{n},\hat{m})\rightarrow\infty$
when $\vec{r}\rightarrow-\hat{n}.$ Therefore, the optimization with
respect to the measurement axis $\hat{m}$ along which $A$ has to
measure is simple since $\mathcal{\mathcal{F}}(\hat{n},\hat{m})$
is a decreasing function of $\hat{n}E\hat{m}^{T}$. Then, since $\hat{n}E\hat{m}^{T}=(\hat{n}E)\cdot\hat{m}$,
the protocol is then optimized when $\hat{m}$ is parallel to $\hat{n}E$,
i.e. when $A$ measures the observable defined by $\hat{m}=\hat{n}_{E}\equiv\hat{n}E/|\hat{n}E|$
; in this case the post measurement state on $B$ is defined by $\vec{r}_{opt}=\hat{n}^{E}E^{T}-\left(\hat{n}^{E}E^{T}\cdot\hat{\beta}\right)\ \hat{\beta}$,
$\mathcal{\mathcal{F}}(\hat{n},\hat{m})$ is minimal and reads

\begin{equation}
\mathcal{F}\equiv\mathcal{F}(\hat{n},\hat{n}^{E})=1-\log(1+|\hat{n}E|).\label{eq: relative entropy maximal payoff}
\end{equation}
Note that $\mathcal{F}(\hat{n},\hat{m})$ is a monotonic function
of $\hat{n}E\hat{m}$, which in the literature is called the ``payoff''
of the protocol (see e.g. \cite{DakicRSP}); correspondingly, the
optimal measurement $n^{E}$ is the same found in the literature and
$\mathcal{F}$ is a monotonic function of the ``optimal payoff''
$|\hat{n}E|$ (for a discussion about different figures of merit see
also \cite{horodeckiRSP} ). The above definition immediately leads
to identify the sub-manifold of \textit{relevant observables} $\Omega_{RO}\subset S^{2}\times S^{2}$
as the set $\Omega_{RO}=\left\{ \left(\hat{n}^{E},\hat{n}\right)|\ \hat{n}\in S^{2}\right\} $.
In order to evaluate the average performance of the protocol we compute
$\left\langle \mathcal{\mathcal{F}}\right\rangle =\left\langle \mathcal{F}(\hat{n},\hat{n}^{E})\right\rangle _{\Omega_{RO}}$
where the average is taken over the submanifold of \textit{relevant
observables $\Omega_{RO}$}; since $\Omega_{RO}\sim S^{2}$, the average
is computed with respect the Haar measure over $S^{2}$. Since $|\hat{n}E|=\sqrt{\hat{n}EE^{T}\hat{n}^{T}}$,
one gets 
\begin{prop}
at fixed $\kappa$, for all states corresponding to a given class
$\mathcal{LU}_{\hat{c}}^{eq}$ defined by $\hat{c}$: $i)$ $\mathcal{F}(\hat{n},\hat{n}^{E})$
is invariant with respect to the action of $\mathcal{LU}$ on $\rho$;
given a state $+\hat{n}$ to be transferred, all states connected
via $\hat{n}=\hat{n}O_{B}$, where $O_{B}$ is the $SO(3)$ representation
of $U_{B}$ such that $U_{A}\otimes U_{B}\in\mathcal{LU}$, have the
same value of $\mathcal{F}(\hat{n},\hat{n}^{E})$; $ii)$ the average
payoff $\left\langle \mathcal{F}\right\rangle _{\Omega_{RO}}$ is
the same for all states corresponding to $\mathcal{LU}_{\hat{c}}^{eq}$. 
\end{prop}
The first property is simply a consequence of the symmetry of the
states i.e., $\hat{n}E\hat{m}^{T}=\hat{n}O_{A}EO_{B}^{T}\hat{m}^{T}=\hat{n}_{A}E\hat{m}_{B}$;
in order to transfer $\hat{n}_{A}$ one has to measure onto $\hat{m}_{B}=\hat{n}O_{A}/|\hat{n}O_{A}|$
with $|\hat{n}O_{A}|=|\hat{n}E|$. The second property is a consequence
of the invariance of the Haar measure with respect to local changes
of bases that realize the given $\mathcal{LU}$. Finally, both $\mathcal{F},\left\langle \mathcal{F}\right\rangle _{\Omega_{RO}}$
are decreasing functions of $\kappa$: the purer the state, the better
the (average) result of the protocol\emph{.} \\
After having identified the relevant observables, one wants to know
what is the benefit of using the correlations present in the state.
By this we mean the following. Suppose one does not use the correlations
present in the state. This can realized if $B$ does not perform the
conditional rotation on his qubit, such that the output of the protocol
is $\vec{r}=\bar{b}=\bar{0}$, corresponding to the identity operator
$\tilde{\varrho}^{B}(\vec{r})=\mathbb{I}_{2}$; in this case $\mathcal{F}$
(\ref{eq: relative entropy maximal payoff}) is independent of $\hat{n}$
and simply reads

\begin{equation}
\mathcal{F}_{\bar{0}}=1	 
\end{equation}
Note that the same result would be obtained if: $i)$ $A$ measures
an observable $\hat{m}$ such that $\hat{n}E\hat{m}=0$ i.e., an observable
that has zero correlations with respect $\hat{n}$; $ii)$ $A$ does
not implement any measurement and always sends the bit $0$ to $B$.
For any desired output $\hat{n}$ a simple way to compare the two
protocols - the one that uses vs the one that does not use the correlations
- is to compare the corresponding probability distributions: $p_{\pm}(\vec{r}_{opt})=(1+|\hat{n}E|)/2$,
i.e., the probability of measuring $\pm\hat{n}$ on $\vec{r}_{opt}$;
and $p_{\pm}(\hat{0}^{E})=1/2$, i.e., the probability of measuring
$\pm\hat{n}$ on $\vec{r}=\hat{0}$. By computing the relative entropy
of the two distributions and with some simple algebra one obtains
\begin{small}
\begin{eqnarray}
\mathcal{D}(\hat{n}^{E}||\hat{0}^{E}) & =\sum_{i=\pm}p_{i}(\vec{r}_{opt})\log_{2}\frac{p_{i}(\vec{r}_{opt})}{p_{i}(\hat{0}^{E})}=\mathcal{I}(\hat{n}^{E},\hat{n})\label{eq: Relative entropy as Gain for MMMS}
\end{eqnarray}
\end{small}
We define $\mathcal{G}(\hat{n},\hat{n}^{E})=\mathcal{D}(\hat{n}^{E}||\bar{0})\equiv\mathcal{I}(\hat{n}^{E},\hat{n})$
as the \textit{gain function} of the protocol\textit{. }The meaning
of the gain stems in the first place from its definition in terms
of relative entropy: the higher $\mathcal{G}$, the higher the statistical
distinguishability between the probability distributions $p_{\pm}(\vec{r}_{opt}),\ p_{\pm}(\hat{0}^{E})$
obtained by using or not using the correlations; in particular if
$p_{\pm}(\vec{r}_{opt})=p_{\pm}(\hat{0}^{E})$ then $\mathcal{G}=0$
and there is no profit in using the correlations. Eq. (\ref{eq: Relative entropy as Gain for MMMS})
establishes a clear connection between\textit{ the gain one gets in
using the correlations in the state and the correlations between the
relevant observables} \textit{$\Omega_{RO}$} as measured by the mutual
information $\mathcal{I}(\hat{n}^{E},\hat{n})$. This is one of the
main results of our analysis: the correlations pertaining to the RSP
for a given state $\rho_{AB}$ are those among the available ones
that are relevant for the protocol. Thus, if one evaluates the average
gain $\left\langle \mathcal{G}\right\rangle _{\Omega_{RO}}=\left\langle \mathcal{I}\right\rangle _{\Omega_{RO}}$,
where the average is taken over the set of relevant observables $\Omega_{RO}$,
one immediately has a measure of correlations tailored to the overall
protocol. The next proposition shows that the gain enjoys the same
properties as the figure-of-merit $\mathcal{F}$.
\begin{prop}
at fixed $\kappa$, for all states corresponding to a given class
$\mathcal{LU}_{\hat{c}}^{eq}$: $i)$ $\mathcal{G}(\hat{n},\hat{n}^{E})$
is invariant with respect to the action of any $U_{A}\otimes U_{B}\in\mathcal{LU}$;
in particular, $\forall\hat{n}$ all observables $\hat{n}O_{B}$,
where $O_{B}$ is the $SO(3)$ representation of $U_{B}$ such that
$U_{A}\otimes U_{B}\in\mathcal{LU}$, have the same value of $\mathcal{G}(\hat{n},\hat{n}^{E})$;
$ii)$ the average gain $\left\langle \mathcal{G}\right\rangle _{RO}$
is the same for all states corresponding to $\mathcal{LU}_{\hat{c}}^{eq}$ 
\end{prop}
The proof simply follows from the proof of Proposition 3 and the fact
that both $\mathcal{F}(\hat{n},\hat{n}^{E})$ and $\mathcal{G}(\hat{n},\hat{n}^{E})$
only depend on $|\hat{n}E|$. One has that both $\mathcal{G},\left\langle \mathcal{G}\right\rangle _{\Omega_{RO}}$
are increasing functions of $\kappa$ i.e., \textit{\emph{t}}he purer
the state the higher (in average) the correlations between the relevant
observables and the higher the profit one gets in using the correlations.
Finally, due to the above definitions of $\mathcal{F}$ and $\mathcal{G}$
- and thanks to the connection between correlation and coherence previously
found (\ref{eq: coherence vs MI}) - one has that for MMMS
\begin{prop}
Given the desired output $\hat{n}$ and the measurement $\hat{m}$
on A: the optimization of the RSP protocol is equivalent to maximizing
the correlations between the observables $\hat{n}\cdot\vec{\sigma}_{A}$
and $\hat{m}\cdot\vec{\sigma}_{B}$ or equivalently to minimizing
coherence with respect to $\rho_{AB}$ of the product bases defined
by $\hat{n}$ and $\hat{m}$. 
\end{prop}
Our scheme therefore allows one to neatly distinguish what is the
relevant resource that matters for the optimization of the RSP protocol
and to and quantify it in the form of the average gain $\left\langle \mathcal{G}\right\rangle _{\Omega_{RO}}$.
In particular, our scheme allows one to identify the \textit{RSP as
a protocol that is based on correlations rather than on coherence.}

\subsubsection*{$\left\langle \mathcal{F}\right\rangle _{\Omega_{RO}}$ and $\left\langle \mathcal{G}\right\rangle _{\Omega_{RO}}$
for $\mathcal{LU}_{\hat{c}}^{eq}$ states\label{sub:Average-pay-off-and-Gain for LU MMMS states}}

We now specify the previous results to some of the classes of states
$\mathcal{LU}_{\hat{c}}^{eq}$ defined in the previous section and
we discuss their properties. Both $\left\langle \mathcal{F}\right\rangle _{\Omega_{RO}}$
and $\left\langle \mathcal{G}\right\rangle _{\Omega_{RO}}$ can be
analytically evaluated in simple cases i.e., for the classes $\mathcal{LU}_{3iso}$
and $\mathcal{LU}_{1iso}$ . For $\rho_{3iso}$ states $\kappa\in\left[0,\sqrt{3}\right]$
and one has

\begin{equation}
\left\langle \mathcal{F}_{3iso}\right\rangle _{\Omega_{RO}}=1-\log_{2}(1+\kappa/\sqrt{3})\label{eq: average F Isotropic 3iso MMMS}
\end{equation}
\begin{eqnarray}
\left\langle \mathcal{G}_{3iso}\right\rangle _{\Omega_{RO}} & = & \left\langle \mathcal{I}_{3iso}\right\rangle _{\Omega_{RO}}=\nonumber \\
 & = & \frac{1}{2}\big((1+\kappa/\sqrt{3})\log_{2}(1+\kappa/\sqrt{3})\nonumber \\
 & + & (1-\kappa/\sqrt{3})\log_{2}(1-\kappa/\sqrt{3}\big)\label{eq: average Gain Isotropic 3iso MMMs}
\end{eqnarray}
For $\rho_{2iso}^{0}$ (the so-called ``classical states'') $\kappa\in\left[0,1\right]$
and 
\begin{equation}
\left\langle \mathcal{F}_{2iso}^{0}\right\rangle _{\Omega_{RO}}=1-\frac{(1+\kappa)\ln(1+\kappa)-\kappa}{\kappa\ln2}
\end{equation}
\begin{eqnarray}
\left\langle \mathcal{G}_{2iso}^{0}\right\rangle _{\Omega_{RO}} & = & \left\langle \mathcal{I}_{2iso}^{0}\right\rangle _{\Omega_{RO}}= \frac{(1+\kappa)^{2}\ln(1+\kappa)}{4\kappa\ln2} - \nonumber \\
 & - & \frac{(1-\kappa)^{2}\ln(1-\kappa)+2\kappa}{4\kappa\ln2}\label{eq: Aveage_Gain_MMMS_2iso0}
\end{eqnarray}
The above functions are important since the classes of states $\mathcal{LU}_{3iso}$
and $\mathcal{LU}_{2iso}^{0}$ are \textit{extremal }in the sense
specified by the following proposition, that holds\textit{ for all
two-qubit states}, as we shall see when we discuss non-MMMS.
\begin{prop}
i) For purity $\kappa\le1$, \textup{$\rho_{3iso}$} states attain
the minimum of both $\left\langle \mathcal{F}\right\rangle _{\Omega_{RO}}$
and $\left\langle \mathcal{G}\right\rangle _{\Omega_{RO}}$ while
the maximum is attained by the class of $\rho_{2iso}^{0}$ states;
ii) For $1\le\kappa\le\sqrt{3}$, the minimum of both $\left\langle \mathcal{F}\right\rangle _{\Omega_{RO}}$
and $\left\langle \mathcal{G}\right\rangle _{\Omega_{RO}}$ is attained
by \textup{$\rho_{3iso}$} while the maxima are found at the intersection
between the sphere of radius $\kappa$ and the tetrahedron $\mathcal{T}$. 
\end{prop}
The proof is given in Appendix \ref{sec:Appendix-B}. Proposition
6 identifies the classes of states i.e., $\rho_{3iso}$ that allow
to obtain, at fixed $\kappa$ the best performance both in terms of
$\mathcal{F}$ and resources needed in RSP. From Proposition 6 follows
that $\rho_{3iso}$ states are those that \textit{for a fixed amount
of average relevant resources $\left\langle \mathcal{G}\right\rangle _{\Omega_{RO}}$
give the best performance} i.e., the smallest $\left\langle \mathcal{F}\right\rangle _{\Omega_{RO}}$.
On the other hand, if \textit{one fixes the value of $\left\langle \mathcal{F}\right\rangle _{\Omega_{RO}}$,
}$\rho_{3iso}$ states are those\textit{ that require the least amount
of resources} to obtain the same performance. The previous statements
are exemplified in Fig. 2.

\begin{figure}
\includegraphics[scale=0.6]{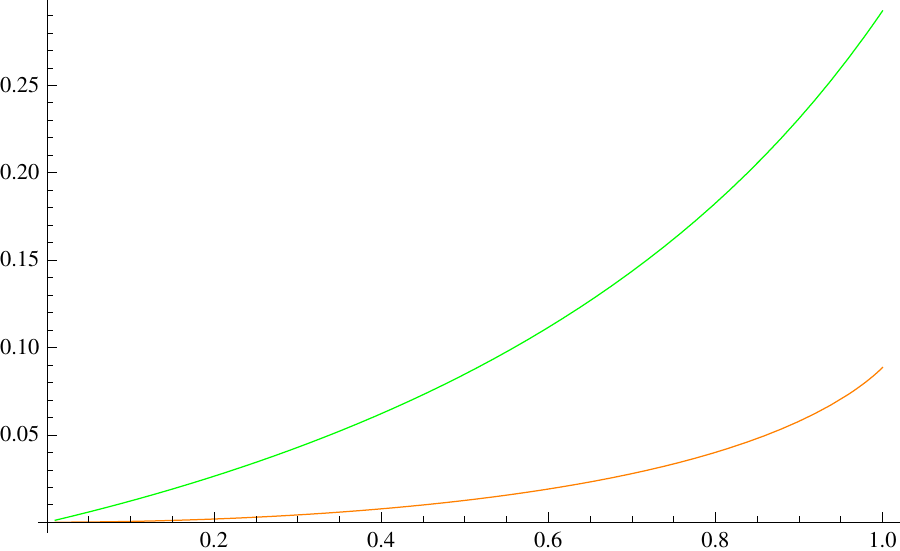}
\protect\caption{(orange) Relative difference between the average gains $\left\langle \mathcal{G}\right\rangle _{\Omega_{RO}}$
for $\rho_{2iso}^{0}$ and $\rho_{3iso}$; (green) Relative difference
between the average figures-of-merit $\left\langle \mathcal{F}\right\rangle _{\Omega_{RO}}$
for $\rho_{2iso}^{0}$ and $\rho_{3iso}$ . At fixed purity symmetry
properties entail differences of up to$8\%$ for the gains and up
to $25\%$ for the figures-of-merit}
\label{Fig. 3}
\end{figure}

These results can be understood since in case of $\rho_{3iso}$ states
the output state of the protocol $\hat{r}=E^{T}\hat{n}^{E}\propto\hat{n}$
i.e. it is always orthogonal to the given $\hat{\beta}$ and parallel
to the desired state output state $+\hat{n}$; therefore $\Omega_{RO}\sim S^{2}\equiv\Omega_{Max}$
i.e., the manifold of relevant observables coincide with the manifold
of maximally correlated observables. For non-isotropic states this
is in no longer true except for a subset of states. For example for
$\rho_{2iso}^{\epsilon}$ this is true iff $\hat{\beta}=\hat{z}$
i.e., for the manifold $S^{1}\sim\Omega_{Max}\subset\Omega_{RO}$
of maximally correlated states, while for $\rho_{2iso}^{0}$ there
is a single pair of observables $\left(\hat{m}=\hat{z},\hat{n}=\hat{z}\right)$.
Therefore for non-isotropic states and for a general desired output
$\hat{n}$, $\hat{r}\nparallel\hat{n}$: therefore in order to obtain
the same value of $|\hat{n}E|$, and therefore the same $\mathcal{F}$,
non-isotropic states must have a higher value of $\kappa$: they must
be purer and employ more resources, in terms of correlations between
the relevant observables, than the isotropic ones\textit{. }\\
Our results can be summarized in the following way: \textit{for a
given state $\rho_{AB}$ the actual resources used in RSP are on one
hand the purity, that determines the amount correlations between relevant
observables as measured by $\left\langle \mathcal{G}\right\rangle _{\Omega_{RO}}$, 
and on the other hand the way (symmetry) in which the correlations
are distributed}. Figure (\ref{Fig. 3}) exemplifies the role of symmetry
at fixed purity by showing the relative differences $\delta\mathcal{G}=\left(\left\langle \mathcal{G}_{2iso}^{0}\right\rangle _{\Omega_{RO}}-\left\langle \mathcal{G}_{3iso}\right\rangle _{\Omega_{RO}}\right)/\left\langle \mathcal{G}_{3iso}\right\rangle _{\Omega_{RO}}$
and $\delta\mathcal{\mathcal{F}}=\left(\left\langle \mathcal{F}_{2iso}^{0}\right\rangle _{\Omega_{RO}}-\left\langle \mathcal{F}_{3iso}\right\rangle _{\Omega_{RO}}\right)/\left\langle \mathcal{F}_{3iso}\right\rangle _{\Omega_{RO}}$
as function of $\kappa$; while the gains differ of at most $8\%$,
the corresponding figures of merit differ up to $25\%$. At fixed
purity symmetry properties entail large differences in the figures
of merit.

 Our treatment of the RSP explains the results presented the literature
from a quite different point of view. For example, in \cite{DakicRSP},
the average performance at given axis $\hat{\beta}$ is expressed
in terms of $\left\langle |E\hat{n}|^{2}\right\rangle _{S(\hat{\beta})}$,
where $S(\hat{\beta})$ is the circle on the Bloch sphere orthogonal
to $\hat{\beta}$ . If one minimizes this average performance with
respect to the choice of $\hat{\beta}$ one has that $\min_{\hat{\beta}}\left\langle |E\hat{n}|^{2}\right\rangle _{S(\hat{\beta})}=(c_{1}^{2}+c_{2}^{2})/2$,
where $|c_{1}|,|c_{2}|$ are the minimal singular values of the correlation
tensor $E$. Therefore, the worst case is given by $\rho_{2iso}^{0}$
states for which $\min_{\hat{\beta}}\left\langle |E\hat{n}|^{2}\right\rangle _{S(\hat{\beta})}=0$.
In our language this simply follows from the symmetry properties of
such states that implies the existence of a circle $S(\hat{\beta})$
of relevant observables that are in fact uncorrelated; therefore on
this circle $\mathcal{G}=0$ and $\mathcal{F}$ is maximal (worst).$ $\\

\subsection{States with non maximally mixed marginals $\bar{a},\bar{b}\protect\neq0$\label{sub: non-MMMS pay-off, gain, relevant observables}}

We now pass to analyze the states with non maximally mixed marginals.
In this case the state prepared by the protocol is $\boldsymbol{r}=E^{T}\hat{m}+\big(\mathbf{b}-E^{T}\hat{m}\big)\cdot\hat{\beta}\ \hat{\beta}$.
Since the state to be transferred is orthogonal to $\hat{\beta}$,
$\hat{n}\cdot\vec{r}=\hat{n}E\hat{m}^{T}$ and therefore the performance
is still given by Eq. (\ref{eq: relative entropy as pay off}) and
$A$ can maximize it (\ref{eq: relative entropy maximal payoff})
by performing a measurement defined by the same observable $\hat{m}=\hat{n}^{E}=\hat{n}E/|\hat{n}E|$.
The protocol therefore relies\textit{ on the correlations of the MMMS}
\textit{that can be obtained by setting $\bar{a}=\bar{0},\bar{b}=\bar{0}$
i.e., $\rho_{AB}(\bar{a}=\bar{0},\bar{b}=\bar{0})$}. Therefore the
condition that leads to the choice of the optimal measurement $\hat{m}=\hat{n}^{E}$
is equivalent to maximizing the correlations between $\hat{m}$ and
$\hat{n}$ that are present in \textit{$\rho_{AB}(\bar{a}=\bar{0},\bar{b}=\bar{0})$
}rather than in\textit{ $\rho_{AB}$.}\\
As for the evaluation of the gain, for non-MMMS states, one is led
to compare two different situations. In the first case the procedure
that makes use of the correlation is the same as the one described
for MMMS, and we refer to it as $\mathcal{P}_{\bar{b}}^{U}$; correspondingly
the figure-of-merit in the optimal case is again (\ref{eq: relative entropy maximal payoff}).
In the second case, in which correlations are not used, one can implement
a procedure that is based on the polarization properties of $\rho_{B}$.
This procedure, which we call $\mathcal{P}_{\bar{b}}^{UN}$ for a
reason that will shortly be clear, can be implemented as follows:
if $\hat{n}\cdot\vec{b}>0$ ($\hat{n}\cdot\vec{b}<0$), $A$ always
sends the bit $0(1)$ so that $B$ never (always) rotates its state,
and the post measurement state is correspondingly $\bar{b}\ \left(-\bar{b}\right)$.
With this procedure the probability of measuring $\hat{n}$ on $\vec{b}$
($-\vec{b}$) is $p_{\pm}(\vec{b})=(1\pm|\hat{n}\cdot\vec{b}|)/2$.
Therefore for $\mathcal{P}_{\bar{b}}^{UN}$, the figure-of-merit $\mathcal{F}^{UN}=\mathcal{F}^{UN}(\hat{n},\bar{b})$
can be derived as in (\ref{eq: relative entropy as pay off}) and
it reads

\begin{equation}
\mathcal{F}^{UN}=1-\log_{2}(1+|\hat{n}\cdot\bar{b}|)\label{eq: relative entropy as pay off non MMM}
\end{equation}
Introducing the procedure $\mathcal{P}_{\bar{b}}^{UN}$ allows us
to devise the following optimized protocol $\mathcal{P}_{\bar{b}}^{opt}$
which has a higher efficiency than the original RSP. Indeed, what
now $A$ must do, for any given $\hat{n}$, is to choose whether to
use or not the correlations present in the state, i.e., whether to
use the procedure $\mathcal{P}_{\bar{b}}^{U}$ or $\mathcal{P}_{\bar{b}}^{UN}$.
To this aim, $A$ must compare the figures-of-merit of the two procedures:
whenever $\mathcal{F}^{U}(\hat{n})<\mathcal{F}^{UN}(\hat{n})$ i.e,
whenever $|\hat{n}E|>|\hat{n}\cdot\bar{b}|$, $A$ uses the state's
correlations; otherwise $A$ does not use them and enacts $\mathcal{P}_{\bar{b}}^{UN}$.
Thus, depending on the desired output state, correlations can be \emph{useful}
or \emph{unuseful} for optimizing the overall RSP performance. This
fact leads us to identify as the resources needed for RSP the correlations
that are\textit{ both relevant and useful}. Given\textit{ $\kappa$
and $\bar{b}$}, the \textit{set of ``relevant and useful observables''
i.e., those that provide relevant and useful correlations, is $\Omega_{\kappa,b}^{U}=\left\{ (\hat{n}^{E},\hat{n})|\ |\hat{n}E|>|\hat{n}\cdot\bar{b}|\right\} $}.
The set of relevant observables is therefore given by the disjoint
union $\Omega_{RO}=\Omega_{\kappa,b}^{U}\bigcup\Omega_{\kappa,b}^{UN}$,
where $\Omega_{\kappa,b}^{UN}=\left\{ (\hat{n}^{E},\hat{n})|\ |\hat{n}E|<|\hat{n}\cdot\bar{b}|\right\} $
is the set of relevant but ``unuseful'' observables, since $(\hat{n}^{E},\hat{n})\in\Omega_{\kappa,b}^{UN}$
, $\mathcal{F}^{U}(\hat{n})>\mathcal{F}^{UN}(\hat{n})$\textit{ }.
The overall figure-of-merit of our optimized protocol $\mathcal{P}_{\bar{b}}^{opt}$
can then be written as :

\begin{equation}
\mathcal{F}^{opt}=\mathcal{F}^{U}(\hat{n})\,\chi_{\Omega_{\kappa,b}^{U}}+\mathcal{F}^{UN}(\hat{n})\,\chi_{\Omega_{\kappa,b}^{UN}}\label{eq: F_4_Modified_Protocol}
\end{equation}
where $\chi_{\Omega_{\kappa,b}^{U}}\,\left(\chi_{\Omega_{\kappa,b}^{UN}}\right)$
is the indicator function that identifies the set of useful (unuseful)
observables for given $\kappa,\vec{b}$. We notice that $\mathcal{F}^{opt}$
correctly takes into account the asymmetry of the RSP with respect
to the exchange of the role of $A$ and $B$ and that is manifest
for non-MMMS whenever $\vec{a}\neq\vec{b}$. In order to better understand
which among the relevant correlations are \textit{useful}, one can
simply notice that the condition $|\hat{n}E|>|\hat{n}\cdot\bar{b}|$
is equivalent to requiring the post measurement states $\boldsymbol{r}_{+},\boldsymbol{r}_{-}$
defined in (\ref{eq: RSP post-measurement states}) to satisfy

\begin{equation}
(\hat{n}\cdot\boldsymbol{r}_{+})(\hat{n}\cdot\boldsymbol{r}_{-})\le0.\label{eq: antiparallel projections-1-1}
\end{equation}
In other words, the components of the vectors $\boldsymbol{r}_{+}$
and $\boldsymbol{r}_{-}$ along the direction defined by $+\hat{n}$
should be opposite in verse. Indeed, suppose both components have
the same verse of $+\hat{n}$, for example if $(\hat{n}\cdot\boldsymbol{r}_{+})>0,(\hat{n}\cdot\boldsymbol{r}_{-})>0$;
then $R_{\beta}^{\pi}\boldsymbol{r}_{-}$ contributes to the final
output state $\boldsymbol{r}$ (\ref{eq:conditionalaveragestate-1})
with a component parallel to $-\hat{n}$, which is orthogonal to the
desired output state $\hat{n}$. Therefore, the rotation of $\boldsymbol{r}_{-}$
around the $\hat{\beta}$ axis required by the standard RSP protocol
is detrimental to the performance. With our modified protocol the
latter is given by $\left\langle \mathcal{F}^{opt}\right\rangle _{\Omega_{RO}}$,
that now has two contributions $\left\langle \mathcal{F}^{U}(\hat{n})\,\chi_{\Omega_{\kappa,b}^{U}}\right\rangle _{\Omega_{RO}}$,
and $\left\langle \mathcal{F}^{UN}(\hat{n})\,\chi_{\Omega_{\kappa,b}^{UN}}\right\rangle _{\Omega_{RO}}$.
As for the properties of $\mathcal{F}^{opt}$ and $\left\langle \mathcal{F}^{opt}\right\rangle _{\Omega_{RO}}$one
has:
\begin{prop}
$i)$ for fixed $\vec{b}$, $\mathcal{F}^{opt}$ and $\left\langle \mathcal{F}^{opt}\right\rangle _{\Omega_{RO}}$
are decreasing functions of $\kappa$; $ii)$ for given $\kappa,\vec{b}$,
states that are obtained by the transformations that connect the unit
vectors that belong to a given class $\mathcal{LU}_{\hat{c}}^{eq}$
have the same value of $\left\langle \mathcal{F}^{opt}\right\rangle _{\Omega_{RO}}$
. 
\end{prop}
Proof of Property $7$ can be found in Appendix \ref{sec:Appendix-D}.
The above considerations demonstrate that the \textit{our modified
protocol }$\mathcal{P}_{\bar{b}}^{opt}$,\textit{ that distinguishes
between useful and unuseful correlations, can in general give a better
performance than the standard RSP}. \\
We now turn to the definition of the gain function for non-MMMS states.
The procedure is analogous to the one seen for MMMS, the main differences
being two. On one hand, the two probability distributions we want
to compare are now: $p_{\pm}(\vec{r}_{opt})=(1\pm|\hat{n}E|)/2$ i.e.,
the probability of measuring $\pm\hat{n}$ on $\vec{r}_{opt}$; and
$p_{\pm}(\vec{b})=(1\pm|\hat{n}\cdot\vec{b}|)/2$ i.e., the probability
of measuring $\pm\hat{n}$ on $\vec{r}=\vec{b}(-\vec{b})$ , the latter
being the same probability used for the definition of $\mathcal{F}^{UN}$.
On the other hand, we want to restrict the evaluation of the gain
to the set of useful observables $\Omega_{\kappa,b}^{U}$ i.e., for
the part $\mathcal{P}_{\bar{b}}^{U}$ of the protocol that effectively
makes use of the correlations. We therefore have for $\left(\hat{n}^{E},\hat{n}\right)\in\Omega_{\kappa,b}^{U}$
and after some manipulations
\begin{eqnarray}
\mathcal{D}(\hat{n}^{E},\bar{b}) & = & \sum_{i=\pm}p_{i}(\vec{r}_{opt})\log_{2}\frac{p_{i}(\vec{r}_{opt})}{p_{i}(\vec{b})}=\nonumber \\
 & = & \mathcal{I}(\hat{n}^{E},\hat{n})_{(\hat{a}=\bar{0},\hat{b}=\bar{0})}+\nonumber \\
 & + & \frac{1}{2}(1+|\hat{n}E|)\log(1+|\hat{n}\cdot\bar{b}|)+\nonumber \\
 & + & \frac{1}{2}(1-|\hat{n}E|)\log(1-|\hat{n}\cdot\bar{b}|\big)\label{eq: Relative entropy_Gain_nonMMMS}
\end{eqnarray}
The gain $\mathcal{G}^{U}=\mathcal{D}(\hat{n}^{E},\bar{b})$\textit{
explicitly depends on the correlations} $\mathcal{I}(\hat{n}^{E},\hat{n})_{(\hat{a}=\bar{0},\hat{b}=\bar{0})}$\textit{
between the relevant observables for the corresponding MMMS $\rho_{AB}(\bar{a}=\bar{0},\bar{b}=\bar{0})$}.
Therefore the desired measure of correlations for the modified protocol
is simply given by the mutual information $\mathcal{I}(\hat{n}^{E},\hat{n})_{(\hat{a}=\bar{0},\hat{b}=\bar{0})}$.
This implies that it is the correlations properties of \textit{$\rho_{AB}(\bar{a}=\bar{0},\bar{b}=\bar{0})$}
rather than \textit{$\rho_{AB}$} that matter for the protocol. This
shift of attention from \textit{$\rho_{AB}$} to \textit{$\rho_{AB}(\bar{a}=\bar{0},\bar{b}=\bar{0})$}
is a direct result of our approach. A simple study reveals that $\mathcal{G}^{U}$
is a growing function of $\kappa$ and a decreasing function of $b=|\vec{b}|$
. These properties can be understood by first analyzing the case in
which $\Omega_{\kappa,b}^{U}=\Omega_{RO}$ i.e., all relevant correlations
are useful, and by considering the difference $\Delta\mathcal{F}=\left(\mathcal{F}^{UN}-\mathcal{F}^{U}\right)$.
When $\kappa$ grows $\mathcal{F}^{U}$ decreases and thus $\Delta\mathcal{F}$
i.e., the gap between the performance of the two protocols, grows:
it becomes even more convenient to use the correlations in the protocol.
On the other hand if $b$ grows the opposite happens: it is $\mathcal{F}^{UN}$
that decreases and thus $\Delta\mathcal{F}$ becomes smaller. The
behavior of $\mathcal{G}^{U}$ with $\kappa$ and $b$ correctly reproduces
these features. As for the average gain one defines $\left\langle \mathcal{G}^{U}\right\rangle =\left\langle \mathcal{D}(\hat{n}^{E},\bar{b})\chi_{\Omega_{\kappa,b}^{U}}\right\rangle _{\Omega_{RO}}$:
the average is taken over the whole set of relevant observables $\Omega_{RO}\sim S^{2}$
and the integrand is different from zero over the set $\chi_{\Omega_{\kappa,b}^{U}}$
and zero otherwise. When $b$ increases, $\left\langle \mathcal{G}^{U}\right\rangle $
decreases not only due to its functional dependence on $b$ but also
because of the restriction of the domain $\Omega_{\kappa,b}^{U}$
over which it is evaluated. The proper average measure of correlations
for the modified protocol is simply given by the average of the mutual
information $\mathcal{I}(\hat{n}^{E},\hat{n})_{(\hat{a}=\bar{0},\hat{b}=\bar{0})}$
over the set of useful correlations i.e., $\left\langle \mathcal{I}^{U}\right\rangle =\left\langle \mathcal{I}(\hat{n}^{E},\hat{n})_{(\hat{a}=\bar{0},\hat{b}=\bar{0})}\chi_{\Omega_{\kappa,b}^{U}}\right\rangle _{\Omega_{RO}}$$ $.
\\
We conclude this section by analyzing the properties of $\mathcal{G}^{U}$
and $\mathcal{F}^{opt}$ for some relevant classes of non-MMMS.

\subsubsection{Example: pure states}

\begin{figure}
\includegraphics[scale=0.5]{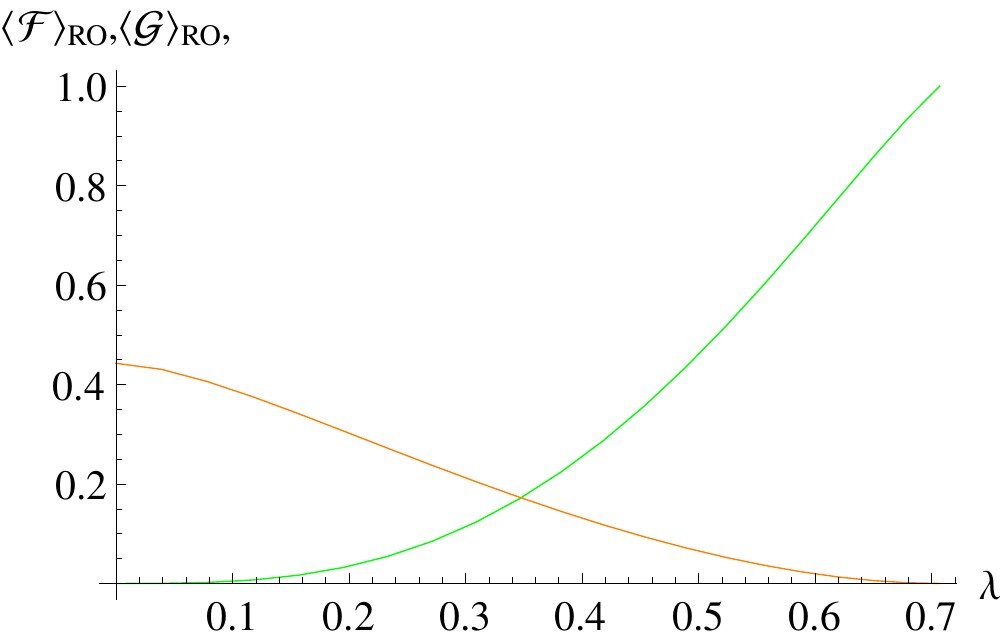}\protect\caption{Average figure of merit $\left\langle \mathcal{F}\right\rangle _{\Omega_{RO}}$
(green) and gain $\left\langle \mathcal{G}^{U}\right\rangle $ (orange)
for pure states as a function of the Schmidt coefficient $\lambda$.}
\label{fig: pure states}
\end{figure}

Thanks to the Schmidt decomposition, the pure states can be written
as $\lambda|00\rangle+\sqrt{1-\lambda^{2}}|11\rangle$ for some choice
of local bases. Therefore, their correlation matrix can be expressed
as $E=\mbox{diag}(2\lambda\sqrt{1-\lambda^{2}},-2\lambda\sqrt{1-\lambda^{2}},1)$
and their local Bloch vectors as $\vec{a}=\vec{b}=(0,0,2\lambda^{2}-1)$
in terms of the single parameter $\lambda$. It is then easy to check
that for pure states $\Omega_{\kappa,b}^{U}=\Omega_{RO}$ i.e., all
relevant observables are useful. In Fig. \ref{fig: pure states}(a)
we plot $\left\langle \mathcal{G}^{U}\right\rangle $ and $\left\langle \mathcal{F}\right\rangle _{\Omega_{RO}}$;
the latter are respectively maximal and minimal for pure Bell states
i.e., given the fixed purity for states maximally isotropic.

\begin{figure}
\includegraphics[scale=0.25]{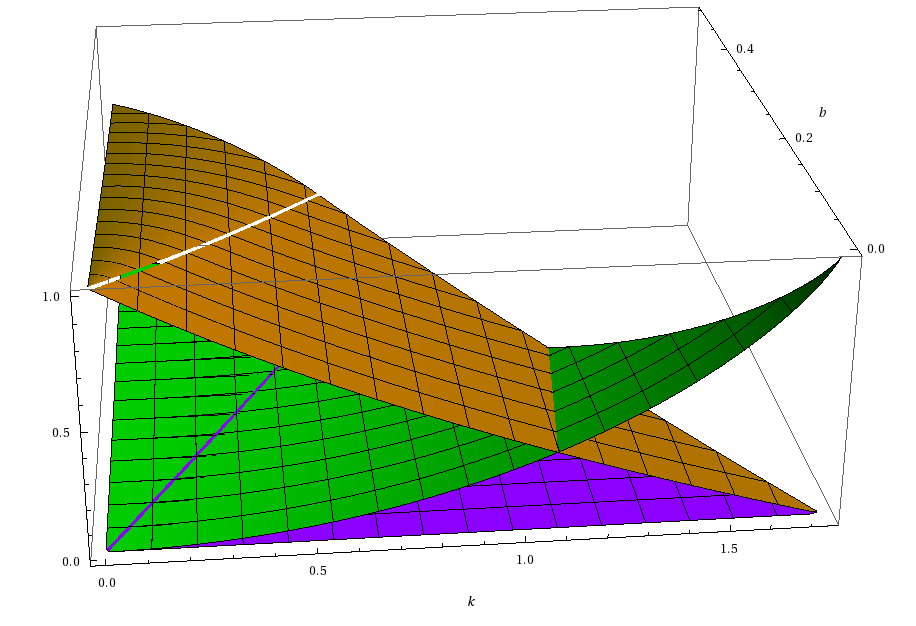}\protect\caption{Figure-of-merit $\left\langle \mathcal{F}^{opt}\right\rangle _{\Omega_{RO}}$
(orange) and gain $\left\langle \mathcal{G}^{U}\right\rangle $ (green)
for isotropic states as a function of $\kappa$ and $b$. The protocol
uses useful correlations only to left of the cuts in the plots. (The
domain of the plot is given by the values of $\kappa$ and $b$ for
which the state is defined) \label{fig:gain=000026F_nonMMMS_Modified_Protocol}}
\end{figure}
\begin{figure}
\includegraphics[scale=0.7]{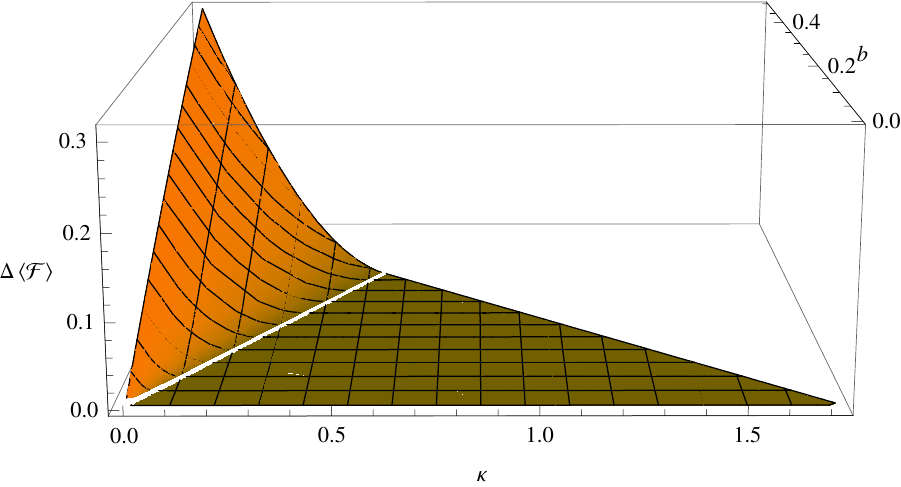}\protect\caption{$\Delta\left\langle \mathcal{F}\right\rangle _{\Omega_{RO}}$: Difference
between the average figure of merit $\left\langle \mathcal{F}\right\rangle _{\Omega_{RO}}$
evaluated for usual RSP that always uses the relevant correlations
{[}Eq. (\ref{eq: average F Isotropic 3iso MMMS}){]} and our modified
protocol based on useful correlations {[}$\left\langle \mathcal{F}_{3iso}^{opt}\right\rangle $
in the text, based on Eq: (\ref{eq: F_4_Modified_Protocol}){]}; when
$\kappa/\sqrt{3}>|\vec{b}|$ one has $\Omega_{RO}\equiv\Omega_{\kappa,b}^{U}$,
the two protocols coincide and $\Delta\left\langle \mathcal{F}\right\rangle _{\Omega_{RO}}=0$;
when $\kappa/\sqrt{3}<|\vec{b}|$ the performance of the modified
better protocol is better i.e., and $\Delta\left\langle \mathcal{F}\right\rangle _{\Omega_{RO}}>0$
(the domain of the plot is given by the values of $\kappa$ and $b$
for which the state is defined) \label{fig:Relative_DeltaF_Normal_vs_Modified}}
\end{figure}

\subsubsection{Example: isotropic case}

As for isotropic case one can first evaluate $\left\langle \mathcal{G}_{3iso}^{U}\right\rangle $
when $\kappa$ and $|\vec{b}|$ are such that $\Omega_{RO}\equiv\Omega_{\kappa,b}^{U}$
i.e., when $\kappa/\sqrt{3}>|\vec{b}\cdot\hat{n}|,\ \forall\hat{n}$
and all relevant correlations are useful. In this case the gain reads
\begin{equation}
\left\langle \mathcal{G}_{3iso}^{U}\right\rangle =\left\langle \mathcal{G}_{3iso}\right\rangle _{\Omega_{RO}}+\left(1-\frac{f\left(1\right)-f(-1)}{6b}\right)/\ln2
\end{equation}
where: $\left\langle \mathcal{G}_{3iso}\right\rangle _{\Omega_{RO}}=\left\langle \mathcal{I}_{3iso}\right\rangle _{\Omega_{RO}}$
is given by (\ref{eq: average Gain Isotropic 3iso MMMs}) i.e., the
result obtained for $|\vec{b}|=0$; while the average of the part
depending on $\vec{b}$ can be written in terms of $f\left(\pm1\right)=(1\pm\ b)\left(3\pm\ \kappa\sqrt{3}\right)\ln\left(1\pm\ b\right)$
and the result depends on $b=|\vec{b}|$ only. In this case $\left\langle \mathcal{F}_{3iso}^{opt}\right\rangle =\left\langle \mathcal{F}_{3iso}^{U}\right\rangle _{\Omega_{RO}}$
and it given by (\ref{eq: average F Isotropic 3iso MMMS}). If now
$\kappa/\sqrt{3}<|\vec{b}|$, $\Omega_{\kappa,b}^{U}\subset\Omega_{RO}$
one has to properly adjust the limits of the integrations in order
to implement both for $\mathcal{G}^{U}$ and $\mathcal{F}^{opt}$
the $\chi_{\Omega_{\kappa,b}^{U}}$ and $\chi_{\Omega_{\kappa,b}^{UN}}$
. The integrations can be carried on analytically and the result is
plotted for the whole set of parameters $\kappa,b$ for which the
state is positive in Figure (\ref{fig:gain=000026F_nonMMMS_Modified_Protocol})
(b). In the figure we show $\left\langle \mathcal{G}_{3iso}^{U}\right\rangle $
and $\left\langle \mathcal{F}_{3iso}^{opt}\right\rangle $; they both
attain their optimal value ($1$ and $0$ respectively) for $\kappa=1$.
The benefit in using our modified protocol can be appreciated in Figure
(\ref{fig:Relative_DeltaF_Normal_vs_Modified}) where we have plotted
the difference between the average figure of merit pertaining to the
usual RSP, given by Eq.: (\ref{eq: average F Isotropic 3iso MMMS})
and $\left\langle \mathcal{F}_{3iso}^{opt}\right\rangle $ for the
optimized protocol $\mathcal{P}_{\bar{b}}^{opt}$. When $\kappa/\sqrt{3}>|\vec{b}|$
one has $\Omega_{RO}\equiv\Omega_{\kappa,b}^{U}$, the two protocols
coincide and they have the same efficiency such that $\Delta\left\langle \mathcal{F}\right\rangle _{\Omega_{RO}}=0$;
when $\kappa/\sqrt{3}<|\vec{b}|$ the optimized protocol has a better
performance , and $\Delta\left\langle \mathcal{F}\right\rangle _{\Omega_{RO}}>0$. 

The results discussed in this Section concern a simple yet paradigmatic
example of quantum communication protocol, RSP; and they show how
the approach introduced allows to define proper protocols-tailored
measures of correlations for both MMMS and non-MMMs states. Furthermore,
the new perspective allows in general to highlight the role of symmetry
in states' correlations distribution (e.g. Proposition 6) and to devise
new optimized protocols that may have better efficiencies. The role
of symmetry is further analyzed from a different perspective in the
following section.

\section{RSP and symmetry\label{sec: RSP and symmetry}}

The main theme of our discussion is the interplay between the two main
resources that characterize the performance of a given quantum protocol:
state purity and correlations symmetry. In particular, we have emphasized the importance of the way the
relevant correlations are distributed in a given state, and how this
property determines the performance of a given protocol. 
In this section
we discuss, in a simplified situation, how the specific kind of symmetry
of a given state $\rho_{AB}$ determines the conditions for the implementation
of the RSP. The original protocol is based on: 
\begin{itemize}
\item the set up of a communication channel, which is realized when $A$
sends part of the state to $B$; 
\item the ability of realizing local measurements along an arbitrary axis
on $A$ side (which is equivalent to the ability of realizing an arbitrary
$SU(2)$ rotation and a measurement along a given fixed axis); 
\item the ability of locally realizing $\pi$ rotations around an arbitrary
axis $\hat{\beta}$ on $B$ side. 
\end{itemize}
The basic RSP requires first the set up of the communication channel,
then after the measurement of the $A$ side and the communication
of the result to $B$, a $\pi$ rotation around a given axis $\hat{\beta}$.
In the following, we analyze the protocol in terms of the resources
needed, in terms of the symmetry of the state and in terms of the
characteristic times of the protocol: $t_{ch},\ t_{\hat{\beta}},\ t_{\hat{n}}$.
We have that $t_{ch}$ is the time in which the channel between $A$
and $B$ is set up; $t_{\hat{\beta}}$ is the time in which the decision
about the axis $\hat{\beta}$ is taken by $A$ and $B$; while $t_{\hat{n}}$
is the time when $A$ gets to know what is the state $\hat{n}$ to
be transferred. Goal of our game is to obtain for states with different
symmetries the same average value of $\mathcal{F}$ (\ref{eq: relative entropy maximal payoff});
this is in general possible but it requires to modify the basic protocol
and to put some constraints in the relations among $t_{ch},\ t_{\hat{\beta}}$,
and $t_{\hat{n}}$. The simplification we adopt is the following:
$A$ sends the $B$ part of the state through a channel that does
not change the state $\rho_{AB}$ initially possessed by $A$ (perfect
channel). This is a quite strong restriction, indeed if the channel
is perfect $A$ could choose to send directly the state $\ket{\hat{n}}$
to $B$. But since we deal with the relation between the performance
of the protocol and the correlations present in a state, the example
allows us to discuss the relation between RSP and symmetry. We focus
on MMMS belonging to different classes and each state will have the
maximum purity allowed by its class, i.e., $\kappa=\max_{\hat{c}\in\mathcal{LU}_{eq}}k_{max}(\hat{c})$.
\\
Suppose now $\rho_{AB}$ belongs to the class $\rho_{3iso}$ , then
$\kappa=\sqrt{3}$ and $\rho_{AB}$ is a pure Bell state. $A$ and
$B$ can then proceed with the usual protocol, and they are free to
choose the three times such that $t_{ch}<t_{\hat{\beta}},\ t_{\hat{n}}$
i.e., $A$ can set up the channel \textit{before knowing $\hat{\beta}$
and $\hat{n}$}. The figure-of-merit of the protocol and the gain
are $\mathcal{F}_{3iso}(\hat{n})=0,\mathcal{G}_{3iso}=1$ for all
$\hat{n}$ and hence also on average. \\
Suppose now $\rho_{AB}$ is such that $\hat{c}=(1/\sqrt{2},1/\sqrt{2},0)$
and $\kappa=1/\sqrt{2}$; the state belongs to the class $\rho_{2iso}^{\epsilon}$,
with $\epsilon=1/\sqrt{2}$, it is separable and its discord is different
from zero. In this case $A$ can do the following: \textit{before}
setting up the channel and \textit{after} she gets to know $\hat{\beta}$,
A rotates $B$'s qubit with a single rotation $U_{\hat{\beta}}$ that
implements on the corresponding Bloch sphere the rotation $\hat{z}\rightarrow\hat{\beta}$;
then $A$ sends the qubit to $B$. And when she knows the desired
$\hat{n}$ , by a proper rotation $U_{xy}$ she rotates her measuring
axes that will lie in the $xy$ plane of her Bloch sphere. The rest
of the protocol is the usual one. It turns out that $\mathcal{F}_{2iso}^{1/\sqrt{2}}(\hat{n})=\mathcal{F}_{3iso}^{\kappa}(\hat{n})>0$
and $\mathcal{G}_{2iso}^{0}=\mathcal{G}_{3iso}^{\kappa}<1$ with $\kappa=\sqrt{3}/2$.
Again the resources used in terms of local rotations i.e., $U_{\hat{\beta}},U_{xy}\sim SU(2)$
are the same as before since in general $U_{\hat{\beta}}$ will be
determined by two real parameters and $U_{xy}$ by a single real parameter
(the angle on the $xy$ circumference). However, in this case it must
be $t_{\hat{\beta}}<t_{ch}<t_{\hat{n}}$. Once again \textit{by using
the same resources and a mixed state one can obtain the same performance
of an isotropic state that for }$\kappa=\sqrt{3}/2$ \textit{is purer
than the state $\rho_{2iso}^{1/\sqrt{2}}$ and it is both entangled
and discordant}.\textit{ }However, symmetry in this case only allows
$A$ to set up the channel \textit{before} knowing $\hat{n}$, but
\textit{after} she gets to know $\hat{\beta}$.\\
Suppose now $A$ is allowed to use the state $\sigma=\left(\left|00\right\rangle \left\langle 00\right|+\left|11\right\rangle \left\langle 11\right|\right)/2$;
the state belongs to the class $\rho_{2iso}^{0}$, $\kappa=1$ and
it is called ``classical'' by some part of the literature \cite{ModiReviewDiscord};
in particular $\sigma$ has zero entanglement and zero discord. $A$
can modify the protocol as follows: instead of using the $SU(2)$
rotations for measuring along different axes, \textit{after A gets
to know both} $\hat{\beta},\hat{n}$ \textit{and before building the
channel} she applies the rotation on the $B$ part of the state such
that $\rho\rightarrow\rho_{\hat{n}}=\left(\left|0\hat{n}\right\rangle \left\langle 0\hat{n}\right|+\left|1,-\hat{n}\right\rangle \left\langle 1,-\hat{n}\right|\right)/2$;
$A$ then sends the second qubit to $B$, implement measurements along
the $\hat{z}$ axis on her qubit and the protocol proceeds as usual.
One has that $\mathcal{F}_{2iso}^{0}(\hat{n})=\mathcal{F}_{3iso}(\hat{n})=0$,
and $\mathcal{G}_{1iso}=\mathcal{G}_{3iso}=1$ for all $\hat{n}$
and on average. The resources used in this case are the same as in
the previous ones ($SU(2)$ rotations on $A$ side, $R_{\pi}(\hat{\beta})$
rotations on $B$ side). Therefore,\textit{ by using the same resources
and a so-called classical mixed state (zero discord and entanglement)
one can obtain the same performance one gets with a pure Bell state}.
The main and relevant difference is that now $t_{\hat{\beta}},\ t_{\hat{n}}<t_{ch}$
i.e., $A$ has to set up the channel \textit{after she gets to know
both $\hat{\beta}$ and $\hat{n}$.} \\
 The bottom line of the above discussion is that, in the described
setup (perfect channel), it is the way the correlations are distributed
among the relevant observables that matters in defining: $i)$ which
kind of freedom one has in realizing the different steps of the protocol
and $ii)$ in which way one has to use the same $SU(2)$ rotations.
The modified protocols for states $\rho_{2iso}^{1/\sqrt{2}},\rho_{2iso}^{0}$
do not change the correlation content of the states; they make use
of the same ability of performing $SU(2)$ rotations as in the original
protocol; the rotations now are used in a way that compensates the
lack of symmetry in the states, in order to reorient the correlation
distribution among the different observables such that the protocol,
as dummy as it may appear, is as efficient as possible with the given
purity. In particular, in the case of the state $\rho_{2iso}^{0}$
the protocol is as efficient as the one that makes use of pure Bell
states. The above results seem to depend on the different symmetries
of the states, rather than the supposed ``quantumness'' or ``classicality''
of the states. Indeed the freedom in the choice of $t_{ch}$ is guaranteed
by the symmetry of the distribution of correlations between the relevant
observables (the ones that are perfectly correlated or anti-correlated).
The states with isotropic correlations allow for a total freedom for
all values of purity, even in absence of entanglement. These states
are always discordant, but here the presence of discord simply records
the presence of a sufficient amount of the ``right symmetry''. \\We
finally note that, in principle, it depends on A's willing or needs
(and on the specific technology at hand) to decide when to set up
the channel. Once the kind of channel to be used is fixed,
the performance of the protocol only depends on the ability of creating 
a state with the highest possible purity and to properly implement the rotations and measurements
needed.\\
Having identified the relevant correlations and their symmetry as
those that determine the performance of RSP, if one relaxes the hypothesis
of a perfect channel, one may argue that the noisy channels that are
optimal are not in general those that preserve entanglement or discord.
On the contrary they are those that preserve the amount of relevant
correlations and the symmetry (isotropy) of the state.

\section{Conclusions\label{sec:Conclusions}}

In this paper we have introduced a new measure of correlations based
on the average classical mutual information $\left\langle \mathcal{I}\right\rangle $
between local von Neumann observables. We have illustrated our measure
focusing on the case of two-qubit systems. To analyze its properties
we defined classes of maximally mixed marginals two-qubit states (MMMS)
with different continuous symmetries. At fixed purity, the states
belonging to each class have the same value of $\left\langle \mathcal{I}\right\rangle $
and their distributions of $\mathcal{I}(\hat{n},\hat{m})$ among the
various observables are isomorphic. At fixed purity, the states that
give the minimum value of $\left\langle \mathcal{I}\right\rangle $
are isotropic states, while those that attain the maximum are those
with a single non zero singular value in their correlation tensor
(the so called ``classical states''). Any pair of local observables
$(\hat{n},\hat{m})$ defines a product basis $\mathcal{B}_{(\hat{n},\hat{m})}$
and we showed for MMMS that the higher $\mathcal{I}(\hat{n},\hat{m})$
the lower the coherence $Coh_{\mathcal{B}_{(\hat{n},\hat{m})}}$ of
the corresponding basis. In other words, the (average) correlations
of MMMS and their (average) coherence are complementary resources:
protocols that require the maximization of $\mathcal{I}(\hat{n},\hat{m})$,
correspondingly require a minimization of $Coh_{\mathcal{B}_{(\hat{n},\hat{m})}}$.
We conjecture that such a distinction\textit{ }may have a general
character and that correlations and coherence may play a complementary
role in quantum information protocols, in the sense that some of them
(or some parts of them) should be based on the maximal amount of correlations
between the relevant observables, and they correspondingly require
the least amount of coherence, while on the contrary others should
be based on the coherence properties of the relevant observables.\textit{}\\
In the rest of the paper, we introduced a general standard scheme
for identifying proper measure of correlations for protocols whose
figure-of-merit $\mathcal{F}(\hat{n},\hat{m})$ explicitly depends
on a given set $\Omega_{RO}$ of pairs of observables $(\hat{n},\hat{m})$
i.e., the set of\textit{ observables} \textit{relevant} for the protocol.
The measure of correlations is obtained by defining a gain function
$\mathcal{G}$ that expresses the benefit in using vs not using the
correlations present in the state $\rho_{AB}$ employed in the protocol.
This perspective has a series of consequences. Indeed, on one hand
the measure of correlations becomes protocol-dependent; on the other
hand the described procedure allows one to derive ``proper'' measures
of correlations in a standard way for each protocol. Ultimately, the
condition of being ``proper'' stems from the explicit connection
one is able to make between the measure of correlations and the figure-of-merit
$\mathcal{F}$. Furthermore, we notice that when a state is sent through
a noisy channels the overall properties of the state are in general
corrupted while, depending on the specific kind of noise, the relevant
correlations may well be preserved. \\
We illustrated our scheme by specializing it to an example of quantum
communication task, remote state preparation (RSP), for which both
discord and entanglement are not able to capture the relevant features
that allow to maximize the performance. In the case of MMMS we introduced
a specific figure-of-merit $\mathcal{F}(\hat{n},\hat{m})$, defined
the set $\Omega_{RO}$ and showed that $\mathcal{G}=\mathcal{I}(\hat{n},\hat{m})$
for $(\hat{n},\hat{m})\in\Omega_{RO}$; therefore the measure of correlations
pertaining to the protocol is just $\mathcal{\left\langle G\right\rangle }_{\Omega_{RO}}=\left\langle \mathcal{I}(\hat{n},\hat{m})\right\rangle _{\Omega_{RO}}$
i.e., the average mutual information between the relevant observables.
The resources involved in the process are the purity of the state and the 
symmetry of the correlations. 
We found that the extremal states are the isotropic ones:
at fixed purity they allow to obtain the optimal value of $\left\langle \mathcal{F}\right\rangle _{\Omega_{RO}}$
with the least amount of $\mathcal{\left\langle G\right\rangle }_{\Omega_{RO}}$
i.e., with the least amount of the resources (correlations) used.
We then extended our scheme to general (non-MMMS) two-qubit states.
The definition of $\left\langle \mathcal{F}\right\rangle _{\Omega_{RO}},\left\langle \mathcal{G}\right\rangle _{\Omega_{RO}}$
parallels that for MMMS, and it shows that the relevant observables
and correlations are those pertaining the state $\rho_{AB}(\vec{a}=\vec{b}=\vec{0})$
i.e, the MMMS obtained from $\rho_{AB}$ by setting the local vectors
$\vec{a},\vec{b}$ to zero. One has that $\left\langle \mathcal{G}\right\rangle _{\Omega_{RO}}$
is a function of $\left\langle \mathcal{I}(\hat{n},\hat{m})\right\rangle _{\Omega_{RO}}^{\vec{a}=\vec{b}=\vec{0}}$
i.e., the average mutual correlation between the relevant observables
evaluated for the state $\rho_{AB}(\vec{a}=\vec{b}=\vec{0})$. Therefore,
$\left\langle \mathcal{I}(\hat{n},\hat{m})\right\rangle _{\Omega_{RO}}^{\vec{a}=\vec{b}=\vec{0}}$
is the desired measure of correlations. Furthermore, for non-MMMS
the study of $\mathcal{F}$ allows one to identify among the relevant
observable the set of those that are indeed \emph{useful} $\Omega_{U}\subset\Omega_{RO}$
and correspondingly to define the $\left\langle \mathcal{F}\right\rangle _{\Omega_{U}},\left\langle \mathcal{G}\right\rangle _{\Omega_{U}}$.
We have shown how to use our approach to devise an optimized protocol
that attains in average better values of $\mathcal{F}$ in a given
range of parameters defining the state $\rho_{AB}$. Our treatment
of RSP allows finding a proper measure of correlations that applies
to all states, identifying classes of states that have the same performance
and discriminating those classes that allow to obtain the best performance
at fixed purity. The optimality of isotropic states has a general
character: the average performance $\left\langle \mathcal{F}\right\rangle _{\Omega_{U}}$
of the protocol is determined by the purity of the state and by the
way (symmetry) in which the useful correlations are distributed.

The idea of analyzing and classifying correlations in terms of classical
mutual information, its average over observables and its symmetries
does not depend on the structure of the set of two-qubit states and
observables. As such, it may be extended to two-qudit and n-qubit
systems, and provide insights into the general structure of quantum
correlations\cite{Maccone}. In addition, our approach to derive protocol
dependent measures of correlations in a standard way may be fruitfully
applied to other relevant protocols. 
\begin{acknowledgments}
We are very grateful to Prof. Matteo G.A. Paris for his careful reading
of the manuscript and his precious suggestions for improvement. We
thank Dr. Giorgio Villosio for his illuminating comments as well as
his enduring hospitality at the Institute for Women and Religion,
Turin (``oblivio c{*}e soli a recta via nos avertere possunt'').\end{acknowledgments}

\section*{Appendix A\label{sec:Appendix-A}}

Given each of the directions $\hat{d}\in\mathcal{LU}_{\hat{c}}^{eq}$
there is always a unique transformation that maps $\hat{c}$ into
$\hat{d}$. These transformations can be seen as orthogonal transformations
in the $\mathbb{R}^{3}$ space of correlation vectors They thus form
a discrete subgroup of $O(3)$ that is isomorphic to $G\sim S_{3}\otimes E_{8}$,
where $S_{3}$ is the symmetric group of order 3, corresponding to
the permutations of three indices, and $E_{8}$ is the elementary
Abelian group of order eight that realizes the changes of signs $s_{i}$
in Eq. (2). This group can be also written as $G\sim S_{4}\otimes Z_{2}$
where $S_{4}$ is the symmetric group of order $4$ and $Z_{2}$ is
the cyclic group of order $2$. The role of the two tensor factors
$S_{4}$ and $Z_{2}$ is best explained by considering the action
of $G$ in the Hilbert space. In the Hilbert space representation,
the transformations of $G$ can implemented by a combination of local
unitary rotations and local spin flips acting on the two-qubit state.
In particular, we have $O_{A}=S_{A}\tilde{O}_{A}$ ,$O_{B}=S_{B}\tilde{O}_{B}$,
where $\tilde{O}_{A},\tilde{O}_{B}\in SO(3)$ and $S_{A},S_{B}\in\left\{ \mathbb{I}_{3},-\mathbb{I}_{3}\right\} $.
The local change of coordinates corresponding to $\tilde{O}_{A},\tilde{O}_{B}$
can always be implemented by means of local unitary operations $U_{A}\otimes U_{B}$
acting on the state. Indeed, it is well known\cite{HorodeckiTetrahedron}
that for any unitary transformation $U\in SU(2)$ there exists a (unique)
rotation $\tilde{O}\in SO(3)$ such that $U\hat{n}\cdot\vec{\sigma}U^{\dagger}=(\tilde{O}\hat{n})\cdot\vec{\sigma}$.
Transformations corresponding to $\tilde{O}_{A},\tilde{O}_{B}$ cannot
change $\det(E)=c_{1}c_{2}c_{3}$ : therefore, acting on $\mbox{diag}(c_{1},c_{2},c_{3})$
they result in permutations of the $c_{i}'s$ and changes of signs
of either zero or two $s_{i}$. These transformations form the subgroup
$S_{4}$of $G$, that can be also interpreted as the symmetry group
of the tetrahedron $\mathcal{T}$, i.e., the group of permutations
of the vertices of $\mathcal{T}$. The other tensor factor group can
be realized as $Z_{2}=\left\{ \mathbb{I}_{3},-\mathbb{I}_{3}\right\} $
where element $-\mathbb{I}_{3}$ realizes the inversion $\hat{c}\rightarrow-\hat{c}$
. The operation represented by the matrix $\mbox{-\ensuremath{\mathbb{I}_{3}}}$
realizes a reflection of one pf the two the qubit's Bloch sphere around
the origin, i.e., a local spin flip of one of the qubits. The spin-flip
cannot be implemented with a unitary operation: in fact, it is anti-unitary
operation \cite{WernerUnot}. If for a given $\kappa$ the vector
$\vec{c}=\kappa\hat{c}$ is admissible (i.e. together with $\vec{a},\vec{b}$
it yields a positive state, then all transformations in $S{}_{4}$,
that can be realized as local unitaries, yield admissible vectors
$\vec{d}=\kappa\hat{d}$. However, the spin flip $-\mathbb{I}$ is
a positive-but-not-completely-positive operation and as such it can
map entangled states into non-positive states. Thus, it may map an
admissible $\vec{c}$ into a non-admissible $\vec{d}$. As proved
in \cite{HorodeckiTetrahedron}, the spin flip is positive only states
such that $\vec{c}\in\mathcal{T}\cap-\mathcal{T}$ . \\

\section*{Appendix B\label{sec:Appendix-B}}

\textit{Proof of Proposition 1.} Since we do not have a general analytical
formula for $\left\langle \mathcal{I}\right\rangle _{(\hat{n},\hat{m})}(\hat{c})$
, we analyze $\left\langle \partial_{\alpha}\left\langle \mathcal{I}\right\rangle _{\hat{n}}(\hat{m})\right\rangle $
and $\left\langle \hat{\partial}_{\beta}\left\langle \mathcal{I}\right\rangle _{\hat{n}}(\hat{m})\right\rangle $
with $\hat{\partial}_{\beta}=\partial_{\beta}/\sin\alpha$ i.e., we
analyze the gradient of $\left\langle \mathcal{I}\right\rangle _{\hat{n}}(\hat{m})$
in spherical coordinates, where $\hat{c}=(\sin\alpha\cos\beta,\sin\alpha\sin\beta,\cos\beta)$.
On has that 
\begin{eqnarray*}
\hat{\partial}_{\beta}\left\langle \mathcal{I}\right\rangle _{\hat{n}}(\hat{m}) & = & \kappa^{2}\left(m_{2}^{2}-m_{1}^{2}\right)\sin\alpha\sin\beta\partial_{R}\left\langle \mathcal{I}\right\rangle _{\hat{n}}(\hat{m})
\end{eqnarray*}
and that $\partial_{R}\left\langle \mathcal{I}\right\rangle _{\hat{n}}(\hat{m})$
is positive $\forall\,R$. Therefore the critical values for $\left\langle \mathcal{I}\right\rangle _{(\hat{n},\hat{m})}(\hat{c})$
are in first place those given by $\partial_{\alpha}R=\partial_{\beta}R/\sin\alpha=0$.
It turns out that for $\hat{c}\in\{\pm\hat{x},\pm\hat{y},\pm\hat{z}\}$
i.e., $\rho_{2iso}^{0}$ state, both derivatives are zero and such
is their average over $\hat{m}$. Furthermore, if one evaluates the
derivatives in correspondence of the isotropic states i.e., $\rho_{3iso}$
one has that $\partial_{R}\left\langle \mathcal{I}\right\rangle _{\hat{n}}(\hat{m})$
is constant with $\hat{m}$ and $\left\langle \partial_{\alpha}R\right\rangle =\left\langle -\sqrt{2}\kappa^{2}(m_{1}^{2}+m_{2}^{2}-2m_{3}^{2})/3\right\rangle =0$
and $\left\langle \hat{\partial}_{\beta}R\right\rangle =\left\langle \kappa^{2}(m_{2}^{2}-m_{1}^{2})\right\rangle =0$.
The evaluation of the average of the Hessian matrix shows that isotropic
states attain minimum and states with single $c_{i}\neq0$ a maximum.
$\rho_{2iso}^{0}$ and $\rho_{3iso}$ states constitute the only extremal
point for $\left\langle \mathcal{I}\right\rangle _{(\hat{n},\hat{m})}(\hat{c})$
and therefore they constitute global maxima and minima. Indeed, the
only other critical points are given by states two $c_{i}$'s equal
and the remaining $c_{j}=0$ i.e., the class of states $\rho_{2iso}^{\epsilon}$
with $\epsilon=1/\sqrt{2}$. In order to show that these are the only
other critical points we focus on the states with with $\alpha=\pi/2,\beta=\pi/4+j\pi,\ j\in\mathbb{Z}$,
since symmetry allow to extend the results to the other elements of
the class $\rho_{2iso}^{\epsilon}$. For the proof it is first sufficient
to show that when $\hat{c}\notin\{\pm\hat{x},\pm\hat{y},\pm\hat{z}\}$
and for non-isotropic states, $\hat{\partial}_{\beta}\left\langle \mathcal{I}\right\rangle _{\hat{n}}(\hat{m})$
has a constant sign for all $\hat{m}$ and therefore $\left\langle \hat{\partial}_{\beta}\left\langle \mathcal{I}\right\rangle _{\hat{n}}(\hat{m})\right\rangle $
cannot be zero, except for $\rho_{2iso}^{1/\sqrt{2}}$ states. To
this aim we express $\hat{m}=(\sin\theta\cos\phi,\sin\theta\sin\phi,\cos\theta)$;
at fixed $\hat{c}$, one has that 
\begin{eqnarray*}
\sin\theta\hat{\partial}_{\beta}\left\langle \mathcal{I}\right\rangle _{\hat{n}}(\hat{m})d\theta d\phi & \propto & -\sin^{3}\theta\cos\left(2\phi\right)\partial_{R}\left\langle \mathcal{I}\right\rangle _{\hat{n}}(\hat{m})d\theta d\phi
\end{eqnarray*}
and therefore the integrand has a constant sign in the integration
over $\theta$. Furthermore, both $\cos\left(2\phi\right)$ and $\partial_{R}\left\langle \mathcal{I}\right\rangle _{\hat{n}}(\hat{m})\doteq\mathcal{I}^{R}(\phi)$
has period $\pi$ as function of $\phi$. The integration for $\phi\in[0,\pi]$
can be replaced by twice the integration for $\phi\in[-\pi/4,3\pi/4]$
and one can show that 
\begin{eqnarray*}
\intop_{-\pi/4}^{3\pi/4}\cos\left(2\phi\right)\mathcal{I}^{R}(\phi) & = & \intop_{-\pi/4}^{\pi/4}\cos\left(2\phi\right)\left(\mathcal{I}^{R}(\phi)-\mathcal{I}^{R}(\phi-\pi/2)\right).
\end{eqnarray*}
Since for $\phi\in[-\pi/4,\pi/4]$ the difference $\mathcal{I}^{R}(\phi)-\mathcal{I}^{R}(\phi-\pi/2)$$ $
has a constant sign (that depends on the sing of $c_{1}^{2}-c_{2}^{2}$.
Therefore, $\hat{\partial}_{\beta}\left\langle \mathcal{I}\right\rangle _{\hat{n}}(\hat{m})$
has a constant sign on the domain of integration and $\left\langle \hat{\partial}_{\beta}\left\langle \mathcal{I}\right\rangle _{\hat{n}}(\hat{m})\right\rangle \neq0$.
The only points in which $\left\langle \hat{\partial}_{\beta}\left\langle \mathcal{I}\right\rangle _{\hat{n}}(\hat{m})\right\rangle =0$
is when $c_{1}^{2}-c_{2}^{2}=0$ i.e., $\beta=\pi/4$. Upon evaluating
$\left\langle \partial_{\alpha}\left\langle \mathcal{I}\right\rangle _{\hat{n}}(\hat{m})\right\rangle $
one finds that it is zero iff $\alpha=\pi/2$. By studying the relative
average of the Hessian one sees that these points are saddle points.
By permutation of the coordinate axes and symmetry arguments, one
can extend the result to the whole set of states $\rho_{2iso}^{1/\sqrt{2}}$
. Since the above arguments are independent on $\kappa$; therefore,
when $\kappa>1$ the domain of $\hat{c}$ shrinks, since some of the
directions define non positive state, and while the minima of $\left\langle \mathcal{I}\right\rangle _{(\hat{n},\hat{m})}(\hat{c})$
remains in correspondence of $\rho_{3iso}$ states, the maxima are
found at the borders of the domain i.e., at the intersection between
the sphere of radius $\kappa$ and the tetrahedron $\mathcal{T}$.

\section*{Appendix C\label{sec:Appendix-C}}

\textit{Proof of Proposition 6. }In order to prove the extremality
of $\rho_{2iso}^{0}$ and $\rho_{3iso}$ for both $\left\langle \mathcal{F}\right\rangle _{\Omega_{RO}}$
and $\left\langle \mathcal{G}\right\rangle _{\Omega_{RO}}$ we use
the same arguments used in Appendix B to proof Proposition 1. Indeed
one can see that, since both $\mathcal{F}$ and $\mathcal{G}$ are
monotonically dependent on $|E\hat{n}|$ $i$) they are monotonically
dependent on $\kappa$; $ii)$ since we do not have an analytical
formula for general $\hat{c}$ , we find the critical points by analyzing
$\left\langle \partial_{\alpha}(R)\right\rangle =\left\langle \partial_{\beta}(R)/\sin\alpha\right\rangle =0$
, with now $R=|E\hat{n}|^{2}=(c_{1}^{2}n_{1}^{2}+c_{2}^{2}n_{2}^{2}+c_{3}^{2}n_{3}^{2})$.
Just as in Appendix B we find that both $\partial_{R}\mathcal{F}$
and $\partial_{R}\mathcal{G}$ are monotonic function of $R$ and
therefore the proof goes along the same line of Appendix B.

\section*{Appendix D\label{sec:Appendix-D}}

Proof. Property $i)$ immediately follows from the following facts:
if $\kappa<\kappa'$ , $\Omega_{\kappa,b}^{U}\subset\Omega_{\kappa',b}^{U}$;
$\mathcal{F}^{U}$ for $(\hat{n},\hat{n}^{E})\in\Omega_{\kappa,b}^{U}\cap\Omega_{\kappa',b}^{U}$
decreases; $\mathcal{F}^{U}<\mathcal{F}^{UN}$ for $(\hat{n},\hat{n}^{E})\in\Omega_{\kappa',b}^{U}\backslash\Omega_{\kappa,b}^{U}$.
Property $ii)$ follows from the following facts: for all transformations
$O_{A},O_{B}$ that maps $\hat{c}\rightarrow\hat{d}$ with $\hat{c},\hat{d}\in\mathcal{LU}_{\hat{c}}^{eq}$
, the sets $\Omega_{\kappa,b}^{U},\Omega_{\kappa,b}^{UN}$ are mapped
into the sets $\Omega_{\kappa,b'}^{U},\Omega_{\kappa,b'}^{UN}$, where
$\vec{b}'=O_{B}\vec{b}$; the result follows from the fact that such
transformations leave the Haar measure invariant. \\

\end{document}